\documentclass[aps,prd,10pt,nofootinbib,twocolumn,eqsecnum,showpacs,showkeys,superscriptaddress,preprintnumbers,altaffilletter]{revtex4-2}
\usepackage{graphicx}
\usepackage{amsmath}
\usepackage{mathtools}
\usepackage{multirow}
\usepackage{hyperref}
\usepackage{soul}
\usepackage{dcolumn}
\usepackage{amssymb}
\usepackage{amsfonts}
\usepackage{amsbsy}
\usepackage{color}
\usepackage{rotating}
\usepackage[english]{babel}
\usepackage{float}
\usepackage{cleveref}
\usepackage{aas_macros}
\usepackage{diagbox}
\usepackage{hhline}
\usepackage{enumerate}

\usepackage{standalone}
\usepackage{import}
\usepackage{tikz} 
\usepackage{colortbl}
\tikzset{
  pics/carc/.style args={#1:#2:#3}{
    code={
      \draw[pic actions] (#1:#3) arc(#1:#2:#3);
    }
  }
}
\usepackage{booktabs}

\newcolumntype{x}[1]{>{\centering\arraybackslash\hspace{0pt}}p{#1}}

\newcommand{\tD}{\tau_D}
\newcommand{\tM}{\tau_M}
\newcommand{\rSch}{R_\mathrm{Sch}}
\newcommand{\dSch}{D_\mathrm{Sch}}
\newcommand{\Gpc}{\mathrm{Gpc}}

\newcommand{\dcc}{P2400321}
\newcommand{\tds}{ET-0474A-24}

\begin{document}

\renewcommand{\tabcolsep}{1.5mm}
\renewcommand{\arraystretch}{1.5}

\title{Invariance transformations in wave-optics lensing: \\
implications for gravitational-wave astrophysics and cosmology}
\date{\today}

\author{Anson Chen}
\email{anson.chen@ligo.org}
\affiliation{Department of Physics and Astronomy, Queen Mary University of London, Mile End Road, London, E1 4NS, United Kingdom}

\author{Paolo Cremonese}
\email{cremonesep25@gmail.com}
\affiliation{Departament de F\`isica, Universitat de les Illes Balears, IAC3–IEEC, E-07122 Palma, Spain}

\author{Jose Mar\'ia Ezquiaga}
%\email{jose.ezquiaga@nbi.ku.dk}
\affiliation{Niels Bohr International Academy, Niels Bohr Institute, Blegdamsvej 17, DK-2100 Copenhagen, Denmark}

\author{David Keitel}
%\email{}
\affiliation{Departament de F\`isica, Universitat de les Illes Balears, IAC3–IEEC, E-07122 Palma, Spain}

\begin{abstract}
Gravitational lensing offers unique opportunities to learn about the astrophysical origin of distant sources, the abundance of intervening objects acting as lenses, and gravity and cosmology in general. 
However, all this information can only be retrieved as long as one can disentangle each effect from the finite number of observables. 
In the geometric optics regime, typical of electromagnetic radiation, when the wavelength of the lensed signal is small compared to the size of the lens, there are invariance transformations that change the mass of the lens and the source-lens configuration but leave the observables unchanged.  
Neglecting this ``mass-sheet degeneracy'' can lead to biased lens parameters or unrealistic low uncertainties, which could then transfer to an incorrect cosmography study. 
This might be different for gravitational waves as their long wavelengths can be comparable to the lens size and lensing enters into the wave-optics limit. 
We explore the existence of invariance transformations in the wave-optics regime of gravitational-wave lensing, extending previous work and examining the implications for astrophysical and cosmological studies. 
We study these invariance transformations using three different methods of increasing level of complexity: template mismatch, Fisher Matrix, and Bayesian parameter estimation. 
We find that, for a sufficiently loud signal, the degeneracy is partially broken and the lens and cosmological parameters, e.g. $H_0$, can be retrieved independently and unbiased.
In current ground-based detectors, though, considering also population studies, a strong constraint on these parameters seems quite remote and the prevailing degeneracy implies a larger uncertainty in the lens model reconstruction.
However, with better sensitivity of the third-generation ground-based detectors, a meaningful constraint on $H_0$ is possible to obtain.
\end{abstract}

\maketitle

\section{Introduction}
\label{sec:intro}

Gravitational lensing of gravitational waves (GWs) occurs when the signal passes close to a massive object that deflects it. This effect could result in multiple images of the same event, magnifications, interference patterns, and delays w.r.t. the unlensed signal. 
With the advancement of the fourth observing run (O4) of the network of LIGO (with the Livingston and Hanford detectors \cite{LIGOScientific:2014pky}), Virgo \cite{Acernese_2015}, and KAGRA \cite{Akutsu:2021, Aso:2013, Somiya_2012} (LVK) and in future runs \cite{KAGRA:2013rdx}, the number of GW detections will grow further, and with it the probability of detecting a lensed event. 
For this reason, in upcoming observing runs, it will be crucial to be able to account for the potential effects of lensing on observed signals. 
Neglecting it could introduce biases into our findings regarding the properties of sources \cite{Cao:2014, Kim:2023viu, Shan:2023qvd}, as lensing magnification, for example, alters the inferred luminosity distance. 
Furthermore, under strong lensing conditions, distortions in the lensed signal may occur, particularly when higher modes, precession, or eccentricity are significant factors \cite{Ezquiaga:2020gdt}.
In the wave-optics regime (i.e., when the signal size is comparable to the size of the lens), the modulation of the signal could mimic some feature of the source, like, for example, the eccentricity, as well as biasing other parameters \cite{Mishra:2023ddt}.
To ensure accurate analyses, it will be necessary to incorporate lensed waveforms into the waveform models for parameter estimation \cite{Nakamura:1997sw, Hannuksela:2019kle, Lai:2018rto, Pang:2020qow, Dai:2020tpj}. Alternatively, employing different methods such as Bayesian analysis \cite{Liu:2020par, Lo:2021nae, Janquart:2021qov} or deep learning approaches \cite{Perreault_Levasseur:2017, Singh:2018csp, Kim:2020xkm} may prove beneficial. Lensing effects also warrant consideration in population studies, as they influence inferred redshift and mass distributions by amplifying signals and allowing the detection of events from greater distances \cite{Oguri:2018muv,Xu:2022a}.

The first observation of a lensed GW is still elusive
after searches in the O1 and O2 runs \cite{Hannuksela:2019kle,McIsaac:2020,Li:2019osa, Dai:2020tpj,Liu:2021}, O3a \cite{LIGOScientific:2021izm}, the full O3 run \cite{LIGOScientific:2023bwz}, and an extended O3 follow-up analysis \cite{Janquart:2023mvf}
yielded no confident signatures.
Nonetheless, there are good reasons to be optimistic about future detections, as many different studies forecast $\sim1$ strong lensing detection per year \cite{Li:2018stc, Ng:2018abc, Xu:2022a, Wierda:2021}.

According to \cite{Bondarescu:2023a}, the probability of microlensing (i.e. in the wave-optics regime) is $<20\%$ of all detectable events above the threshold signal-to-noise ratio (SNR)\footnote{
    In the paper, the threshold SNR is set at $\rho=10$.
    The level of the detectability of a lensed event is influenced by the mismatch between the lensed signal and the (best matching) unlensed waveform because a mismatch $>0$ decreases the detected SNR.
} for $M_{det} = 120~M_\odot$ and $<5\%$ for more common events with $M_{det} = 60~M_\odot$ (where $M_{det}$ is the detector-frame mass).

The detection and subsequent study of lensed GWs is a powerful tool for both astrophysical and cosmological studies. 
For example, gravitationally lensed events can be useful because
(i) contrarily to unlensed GWs, they give an improved localisation of the source \cite{Hannuksela:2020xor, Yu:2020};
(ii) the magnification due to the lensing effect extends the range of detectability of GWs to higher redshift (due to magnification) (see e.g. \cite{Xu:2022a}); 
(iii) through the characterisation of the lens, one can study the content and composition of the universe \cite{Lai:2018rto, Diego:2019a, Oguri:2020a, Cremonese:2021ahz};
the study of one or several lensed systems can be exploited for (iv)  precision cosmology \cite{Sereno:2011ty, Liao:2017ioi, Cao:2019kgn, Li:2019osa, Hannuksela:2020xor},
(v) statistical cosmology \cite{Xu:2022a},
and (vi) tests of general relativity \cite{Baker:2017a, Collett:2017a, Fan:2017a, Goyal:2021a, Ezquiaga:2020gdt, Goyal:2023uvm}.

The time delay between multiple images in gravitational lensing has been a long promise to measure the Hubble constant $H_0$ \cite{Refsdal:1964blz}, since the time-delay distance, which is a combination of the angular diameter distances between source, lens, and observer, is proportional to it. 
This method, also called the time-delay cosmography, was first proposed by Refsdal in 1964 \cite{10.1093/mnras/128.4.307} and can be similarly performed with strongly-lensed GWs. For multiple lensed GW signals from compact binary coalescence, the lens model can be inferred from the magnification ratio and the arrival time delay between them, using ground-based detectors \cite{Liao:2017ioi, Hannuksela:2020xor, Wempe:2022zlk}, 
space-based antennas 
\cite{Sereno:2011ty}, or pulsar timing arrays \cite{Cremonese:2019tgb}. 
If accompanied by electromagnetic (EM) counterparts, the host galaxies of LVK GW sources can be pinpointed using the information from galaxy catalogs \cite{Liao:2017ioi, Hannuksela:2020xor}, so that $H_0$ could be measured at high accuracy. 

Unfortunately, the usefulness of studying lensed events has a large caveat that goes under the name of mass-sheet degeneracy (MSD) \cite{Falco85_MSD, Gorenstein88_MSD, Schneider:2013sxa, Schneider:2013wga}.
The MSD relies on the fact that the simultaneous scaling of the lens mass and the source plane leaves the geometrical lensing outputs unchanged. 
Thus, in practice, we have an intrinsic indeterminacy when we analyze observational data to reconstruct the mass (model) of the lens from the image structure. 
This is a well-known problem in electromagnetic (EM) lensing. 

In EM studies, the MSD is always defined in the \textit{geometrical optics} approximation, when the wavelength of the lensed signal is much smaller than the typical size of the lens. In this limit,
the MSD can be broken, or at least attenuated, when: 
(i) the same gravitational object act as lens for multiple sources at different redshifts \cite{Saha:2000kn, Lubini:2013mta, Grillo:2020yvj, Liao:2017ioi}; 
(ii) independent mass estimations of the lens are available, e.g. from spatially resolved kinematic observations of the lens \cite{Schneider:2013sxa, Birrer:2020tax, Ding:2021bxs}; 
(iii) by a proper combination of the previous astrophysical data with cosmological geometrical data \cite{Chen:2020knz}.
Further discussions in strong-lensing degeneracies can also be found in \cite{Wagner:2018rae,Wagner:2018jxp,Wagner:2019azs}.

In the context of gravitational lensing of GWs, given the much lower frequency of the signal, the \textit{wave optics} regime can also be relevant as GW wavelengths can be comparable to the lens size \cite{Takahashi:2016jom, Cremonese:2018cyg, Cremonese:2019tgb, Jung:2017flg, Dai:2018enj, Sun:2019ztn, Liao:2020hnx}. Even though the MSD can be broken in this limit \cite{Cremonese:2021puh}, such a degeneracy can affect lens mass modeling and can highly bias the estimation of the Hubble constant, $H_0$ \cite{gralen.boo, Schneider:2013sxa, Kochanek:2019ruu}, contributing to the systematic error budget. This is the reason why the MSD is carefully analyzed in projects like H0LiCOW \cite{Chen:2019ejq, Wong:2019kwg}, TDCOSMO \cite{Millon:2019slk} and in the Time Delay Lens modeling Challenge \cite{Ding:2018hai, Ding:2020jmg}. 
In GW lensing, this topic was first discussed in \cite{Cremonese:2021puh} and, more recently in the geometric optics regime in \cite{Poon:2024zxn}.

In this work, we perform rigorous checks on the methods to break the MSD in the wave optics limit and analyze its effects in parameter estimation for the lens model. We further demonstrate the implications of the more general invariance transformations in gravitational lensing to cosmology. 
This paper is organized as follows: 
in Section~\ref{sec:lensing} we introduce the basis of gravitational lensing;
what the invariance transformation in GW lensing are, their role in the different optical regimes and how we study them in different lens models is presented in Section~\ref{sec:transformations};
in Section~\ref{sec:methods} we explain the methods through which we study the MSD;
Section~\ref{sec:implications_astro} and Section~\ref{sec:implications_cosmo} show the implication of the MSD for astrophysical and cosmological studies, respectively;
and in Section~\ref{sec:conclusions} we draw our conclusions.

\section{Gravitational lensing}
\label{sec:lensing}

\subsection{General phenomenology and formalism}

In its essence, gravitational lensing accounts for solving the wave propagation in a curved background \mbox{$\Box h_A + \mathcal{R}_A^Bh_B=0$}, where $A$ indicates a generic wave polarization of the GW strain $h$ and $\mathcal{R}_A^B$ schematically represents the relevant contraction of the background curvature (Riemann tensor) with the given polarization. 
In the weak-field limit, where the gravitational potential is small \mbox{$\Phi\ll1$},
the background metric simplifies to $ds^2\approx -(1+2\Phi)dt^2+(1-2\Phi)d\vec{x}^2$, and the curvature term is subdominant. 
Deflection angles and polarization rotations are small in this limit. 

Therefore, under these approximations, the problem simplifies to that of solving a scalar wave equation on a medium, i.e. a Helmholtz equation (see e.g. \cite{Nakamura:1999uwi} for a review):
\begin{equation}
    \nabla^2 h - (1 - 4\Phi)\partial_t^2 h =0\,. 
\end{equation}
This equation can be solved in Fourier space in terms of a Kirchhoff diffraction integral over the angular frequency $\omega$:
\begin{equation}
    h_L(t,\vec{\theta}_S)= \int\frac{\mathrm{d}\omega}{2\pi}e^{i\omega t}F(\omega,\vec{\theta}_S)h_\omega\,,
\end{equation}
where the whole wave-propagation problem is replaced by solving the amplification factor (in the thin-lens approximation)
\begin{equation}
    F(\omega,\vec{\theta}_S) = \frac{D_LD_S}{cD_{LS}}\frac{\omega}{2\pi i}\int \mathrm{d}^2\theta~\exp[i\omega t_d(\vec{\theta},\vec{\theta}_S)]\,.
\end{equation}
Here $\vec\theta_S$ is the angular position of the source, which with lensing is in general different from the image position $\vec\theta$. $D_L$, $D_S$ and $D_{LS}$ are the angular diameter distances from the observer to the lens, to the source, and between the source and the lens, respectively.  

For later convenience, we also introduce the time scale associated with the angular diameter distances \mbox{$\tD\equiv D_LD_S/cD_{LS}$}. 
The time delay $t_d$ is determined by 
\begin{equation}
    t_d(\vec{\theta},\vec{\theta}_S)\approx \frac{\tD}{2}|\vec{\theta}-\vec{\theta}_S|^2 + t_\Phi\,,
    \label{eq:time_delay}
\end{equation}
where the first term corresponds to the geometric time delay and the second one to the Shapiro delay due to traversing a lensed path $s$ through the gravitational potential $\Phi$
\begin{equation}
    t_\Phi \approx -\frac{2}{c^3}\int \Phi\mathrm{d}s\,.
\end{equation}

Note that the normalization of the amplification factor is such that in the absence of lensing, $t_d=\tD |\vec{\theta}-\vec{\theta}_S|^2/2$, then \mbox{$F=1$}. 
Lensing typically occurs at a reference angular scale, $\theta_*$. If we redefine the coordinates appropriately
\begin{equation}
\vec{x} \equiv \vec\theta / \theta_*\,,\quad \vec{y} \equiv \vec\theta_S / \theta_*\,, \quad w\equiv \tD \theta_*^2 \omega\,,
\end{equation}
then the amplification factor takes a more compact form
\begin{equation}
    F(w,\vec{y}) = \frac{w}{2\pi i}\int \mathrm{d}^2x~\exp[iw T_d(\vec{x},\vec y)]
    \label{eq:F_dimensionless}
\end{equation}
in terms of the dimensionless frequency $w$ and time delay $T_d \equiv t_d / \tD \theta_*^2$. The dimensionless source position $y$ is also referred to as the impact parameter.

The amplification factor $F$ accounts for the time delay associated with all possible lensed paths from the source to the observer, which in general can lead to complicated interference or diffraction patterns. 
There is a limit, however, in which the integral is dominated by its stationary points 
\begin{equation} \label{eq:stationary_points}
    \left.\frac{\partial t_d}{\partial\vec\theta_j}\right|_{\vec\theta=\vec\theta_j}=0\,,
\end{equation}
and lensing is characterized by having distinct images at the locations $\vec\theta_j$. 
For sufficiently strong lenses, multiple images are produced.

In this limit, the amplification factor can be solved using the stationary phase approximation (SPA) leading to
\begin{equation}
    F(\omega,\vec\theta_j)\approx\sum_j \sqrt{|\mu_j|}\exp\left[i\omega t_j-i \mathrm{sign}(\omega)n_j\frac{\pi}{2} \right]\,.
    \label{eq:F_SPA}
\end{equation}
In this limit, lensing is described by a set of $j$ images each arriving at a different time $t_j \equiv t_d(\vec\theta_j)$, with a different magnification $\mu_j\equiv\mu(\vec\theta_j)$, and Morse phase $n_j=0,1,2$ for type I, II, and III images respectively. 
The different Morse phases correspond to the different types of solutions of the lens equation \eqref{eq:stationary_points}: minimum, saddle point, and maximum. 
Note that the term $\mathrm{sign}(\omega)$ is only there to ensure that the lensed signal in the time domain is real.
Moreover, the magnifications are obtained from the determinant of the Hessian matrix evaluated at the stationary points
\begin{equation}
    \mu(\theta_j) = 1/\mathrm{det}(T_{ab}(\theta_j))\,,
\end{equation}
where
\begin{equation}
    T_{ab}\equiv 
    \tD^{-1}\partial^2 t_d/\partial \theta_a\partial\theta_b\,.
\end{equation}
Apart from the phase shift of type II images that can induce waveform distortions on signals with multiple frequency components \cite{Ezquiaga:2020gdt}, lensing is achromatic in the SPA. 

The SPA is satisfied when the arrival time difference between the stationary points is larger than the duration of the signal, i.e. $|\Delta t_{d}\cdot\omega|\gg1$. 
The geometric optics (or eikonal) approximation, determined by the wavelength being smaller than the curvature scale of the lens, $\lambda/\mathcal{R}\ll1$, encompasses the SPA. However, there are source-lens configurations in geometric optics where the SPA is broken, namely at the caustics where the Hessian matrix $T_{ab}$ becomes singular. At those locations, a single, highly magnified image is formed and the full diffraction integral needs to be solved.

The other relevant limit in lensing is that of strong gravity, where $\Phi \sim 1$, and our initial weak-field assumption is broken. In astrophysics, this only occurs very close to extremely dense objects such as black holes, whose characteristic length scale is given by the Schwarzschild radius $\rSch= 2GM/c^2$. 
If we model such a black hole as a point lens, its characteristic lensing scale is the Einstein radius
\begin{equation}
R_E \approx \theta_E D_L = \sqrt{\frac{2(1+z_L)\rSch}{c\tD}}D_L=\sqrt{\frac{\tM}{\tD}}D_L\,,
\end{equation} 
where in the last equality we have introduced the dilated Schwarzschild diameter crossing time \mbox{$\tM = \dSch/c$}, with \mbox{$\dSch=2\rSch$}. 
The ratio of the two scales is then
\begin{equation}
\frac{R_E}{\dSch}=\sqrt{\frac{D_LD_{LS}}{\dSch D_S}}\,.
\end{equation}

In other words, weak-field lensing occurs at scales much larger than the strong gravity regime unless the source is very close to the lens, \mbox{$D_{LS}\ll D_L,D_S$}, as is the case e.g. for the photons near the super-massive black holes imaged by the Event Horizon Telescope \cite{EventHorizonTelescope:2019dse,EventHorizonTelescope:2022wkp}. 
For typical astrophysical lensing situations in which \mbox{$D_{LS}\approx D_L\approx D_S$}, then
\begin{equation}
\frac{R_E}{\dSch}\approx 10^{10}\left(\frac{10 M_\odot}{(1+z_L)M}\right)^{1/2}\left(\frac{D_{LS}}{1\Gpc}\right)^{1/2}\,,
\end{equation}
and strong gravity can be safely neglected for all practical purposes.

\subsection{GW lensing}

So far our treatment of gravitational lensing has been generic to any type of incoming wave. We now specialize to the case of GWs.
GWs from compact binary coalescences (CBCs) are unique compared to other transients and their observing facilities in several aspects: 
\begin{itemize}
\item the signal is coherently detected: amplitude and phase are measured as a function of time,
\item the signal is understood from first principles: general relativity can be used to derive waveform models for CBCs (though numerical simulations and various effective modeling approaches are needed in practice, see \cite{Schmidt:2020ekt}),
\item the wavelength can have astrophysical sizes: the GW frequency is associated to the orbital motion,
\item the detectors nearly continuously monitor all the sky: GWs traverse the Earth from any sky location without being altered,
\item selection effects are well understood and can be quantified through simulations, thus allowing the reconstruction of source and population properties even with incomplete observations.
\end{itemize}
All these characteristics will be relevant for different aspects of gravitational lensing. 
Importantly, GW sky localizations, which are at best of the order of few square degrees in optimistic current observing scenarios \cite{KAGRA:2013rdx}, will be much larger than the typical angular scale of the problem $\theta_*$.

In a strong lensing scenario with $N$ images, the GW lensing observables are the $(N-1)$ time delays between the images $\Delta t_{ij}$, the $(N-1)$ relative magnifications $\mu_i/\mu_j$, and the $N$ Morse phases $n_j$. Given the poor sky localizations, the image locations are not observables, nor is the shape of the source, which is assumed to be point-like. As we will see in the next section, this limits significantly the amount of information that can be obtained about the lens model. 
But in the wave-optics regime, lensing still produces a measurably distorted GW strain time-series (waveform) of the single-lensed image. Given that we can predict this waveform from general relativity to good precision, the whole frequency-dependent amplification factor is in principle an observable. This already hints that in the wave-optics limit of GW lensing there is potentially more information to be reconstructed about the lens model.

\section{Invariance Transformations in gravitational wave lensing}
\label{sec:transformations}

Our ability to learn about astrophysics and cosmology from lensed systems depends on how well we can distinguish the predictions of different models on the observables. 
For this reason, we now focus on studying which sets of variations of the lens system leave the observables unchanged, as those will lead to degeneracies between different astrophysical and cosmological interpretations. 
Such ``invariance transformations'' have been widely studied in the EM spectrum since they constitute a large fraction of the error budget when reconstructing the lens model or inferring the expansion rate of the Universe \cite{Falco85_MSD, Gorenstein88_MSD, Schneider:2013sxa, Schneider:2013wga}. 
They have also been explored in the context of GWs \cite{Cremonese:2021puh, Poon:2024zxn}, and here we will build upon those results. 

\subsection{Multiple-image regime}
\label{sec:multiple_image}

To explain the general working of the invariance transformations, we begin by focusing on the regime of strong lensing, in particular, when multiple images of the original source are formed. 
As we have seen, the different observables (e.g. image positions, relative magnifications, and time delays) can be derived from the general time-delay expression (Eq.~\ref{eq:time_delay}). 
On the other hand, the time delay can be written in terms of the deflection angle $\vec\alpha=\vec\theta-\vec\theta_S$ to obtain:
\begin{equation}
    t_d= \frac{\tD}{2}\cdot|\vec\alpha|^2 - \tD\int \vec\alpha \mathrm{d}\vec\theta\,,
\end{equation}
where we have used the definition of the stationary points in Eq.~\eqref{eq:stationary_points} to rewrite the Shapiro delay in terms of $\vec\alpha$.

Following \cite{Schneider:2013sxa}, we are going to consider three possible modifications of our fiducial lens scenario:
\begin{enumerate}[i)]
    \item increase the surface mass density
    \begin{equation} \label{eq:sigma_lambda}
        \Sigma \to  m\Sigma\,,
    \end{equation}
    \item add a sheet of matter with constant dimensionless surface mass density $\kappa_c$,
    \item change the local expansion rate as parameterized by \mbox{$h_0=H_0/H_0^\mathrm{fid}$}, with the reference value \mbox{$H_0^\mathrm{fid}=70$ km s$^{-1}$ Mpc$^{-1}$}.
\end{enumerate}
In turn, the deflection angle changes as
\begin{equation}
    \vec\alpha\to h_0 m\vec\alpha + \kappa_c \vec\theta\,.
    \label{eq:invariance_transform}
\end{equation}
Similarly, the time delay becomes (for brevity we write $\lambda =h_0 m$):
\begin{equation}\label{eq:t_lambda_1}
    t_d\to \lambda t_d - \frac{\tD}{2}\left[\lambda(1-\lambda)\vec\alpha^2-2\kappa_c\lambda\vec\alpha\vec\theta + \kappa_c(1-\kappa_c)\vec\theta^2\right]\,.
\end{equation}
By taking derivatives of the time-delay surface with respect to the image positions we can obtain again all the observables. 
For example, the Hessian transforms to
\begin{equation}
\begin{split}
    T_{ab}&=\frac{\partial\theta_{S,a}}{\partial\theta_b}=\delta_{ab}-\frac{\partial\alpha_a}{\partial\theta_b} \\
    &\to (1-\kappa_c)\left[\delta_{ab}-\frac{\lambda}{1-\kappa_c}\frac{\partial\alpha_a}{\partial\theta_b}\right]\,,
\end{split}
\end{equation}
where $\delta_{ab}$ is the identity matrix. 
Therefore, in order to keep the relative magnifications unchanged we need to impose 
\begin{equation} \label{eq:lambda_inv}
    \lambda = 1 - \kappa_c\,.
\end{equation}
This same condition leaves the relative image positions $\vec\theta_j-\vec\theta_i$ unchanged. The individual magnifications, time delays between the images and source position change to
\begin{align}
    \mu&\to \mu/\lambda^2\,, \label{eq:mu_lambda}\\
    \Delta t_{ij} &\to \lambda \Delta t_{ij}\,, \label{eq:deltat_lambda}\\
    \vec\theta_S&\to\lambda\vec\theta_S\,.\label{eq:impact_lambda}
\end{align}
Altogether, the amplification factor in the multiple-image regime can be written as
\begin{equation}
\begin{split}
    F_\lambda&(\omega,\vec\theta_j)\approx\frac{\sqrt{|\mu_i|}}{\lambda}e^{i\omega t_{\lambda,i}}\times \\
    &\left(1+\sum_{j\neq i} \sqrt{\frac{|\mu_j|}{|\mu_i|}}\exp\left[i\lambda\omega \Delta t_{ij}-i \mathrm{sign}(\omega)n_j\frac{\pi}{2} \right]\right)\,,
\end{split}
    \label{eq:F_SPA_lambda}
\end{equation}
where we have factorized around a given image $i$ with phase $n_i=0$ and time delay $t_{\lambda}$ given by Eq. (\ref{eq:t_lambda}).

Unless there is additional information, the individual magnifications as well as the source position are not observable. 
Therefore, condition (\ref{eq:lambda_inv}) defines a set of transformations parameterized by a single variable $\lambda$ that leaves all the multiple-image observables unchanged except for the relative arrival times of the images. 
In the following sections, we will use $\lambda$ as a proxy to describe these invariance transformations. 
Because of the combination between increasing the surface mass density $m$ and adding a sheet of dimensionless surface mass density $\kappa$, these invariance transformations are known as \emph{mass-sheet degeneracy} (MSD).

\subsection{Wave-optics regime}

We now explore these invariance transformations in the regime of wave optics. Since the diffraction integral is not analytically solvable in general, we restrict to the simpler case of spherically symmetric lenses. 
In this case, the diffraction integral in Eq.~(\ref{eq:F_dimensionless}) can be simplified to one dimension, integrating over the angular polar coordinate. 
Doing so, one arrives at \cite{Nakamura:1999uwi}
\begin{align}
    F =& -iw\exp\bigg(\frac{1}{2}iw y^2 \bigg) \nonumber\\
    \times& \int_0^{\infty} dx~xJ_0(w xy)\exp\bigg\{iw\bigg[\frac{x^2}{2}-\Psi(x) \bigg] \bigg\},
    \label{eq:F_sphere}
\end{align}
where $J_0$ stands for the Bessel function of the first kind. Under the mass-sheet transformation defined by Eq.~(\ref{eq:lambda_inv}), the time delay is transformed as
\begin{equation}
    t_\lambda(\vec{x},\vec{y}) = \lambda t(\vec{x},\vec{y}) - \frac{\lambda(1-\lambda)}{2} (1+z_L)\tau_D\theta_*^2 |\vec{y}|^2.
    \label{eq:t_lambda}
\end{equation}
The potential of the lens is also transformed as
\begin{equation}
    \Psi_\lambda(\vec{x}) = \lambda \Psi(\vec{x}) + (1-\lambda)\frac{|\vec{x}|^2}{2} \,.
    \label{eq:potential_lambda}
\end{equation}

Substituting the transformed potential and the impact parameter given by Eq. \eqref{eq:potential_lambda} and \eqref{eq:impact_lambda}, the amplification factor for the spherical symmetric lens model is transformed by
\begin{align}
    F_\lambda = & -iw\exp\bigg(\frac{1}{2}iw\lambda^2y^2 \bigg) \nonumber\\
    & \times \int_0^{\infty} dx~xJ_0(\lambda w xy)\exp\bigg\{iw\lambda\bigg[\frac{x^2}{2}-\Psi(x) \bigg] \bigg\}.
    \label{eq:F_lambda}
\end{align}
Defining $\nu_\lambda=-i\lambda w/2$, one can note that this expression is very similar to the previous one:
\begin{equation} \label{eq:F_lambda_sph}
    F_\lambda(w,y) = \frac{e^{(1-\lambda)\nu_\lambda y^2}}{\lambda}F(\lambda w,y)\,.
\end{equation}
One can then conclude that the mass-sheet transformation (\ref{eq:lambda_inv}) rescales the dimensionless frequency, $w\to\lambda w$, and the arrival time ($\nu_\lambda$ defines a complex phase linear in $w$). 

\subsection{Point mass model}

To gain further intuition, we fully specify the lens model to exemplify the effect of the invariance transformations. 
We start with the simplest model: a point mass (PM), where the potential is given by $\Psi(x)=\log x$. The Einstein radius in the PM model is $\theta_* = \sqrt{4GM_{l,z}/c^3\tau_D}$ \cite{Takahashi:2016jom}. 
The amplification factor for the geometric optics regime with the SPA without considering the MSD is given by Eq. \eqref{eq:F_SPA}. 
For a PM lens there are always two images. Their time delay is related to the (dimensionless) source position $y$ by \cite{Takahashi:2016jom}
\begin{equation}
    t_{\pm} = \frac{y^2 + 2 \mp y \sqrt{y^2 +4}}{4}- \log \frac{|y\pm\sqrt{y^2+4}|}{2}.
    \label{eq:time_delay_GO}
\end{equation}
The time interval between the magnified and the demagnified image is just $\Delta t=t_{-}-t_{+}$. In addition, the magnification factor is also given by a simple analytic expression:
\begin{equation}
    \mu_\pm = \frac{1}{2} \pm \frac{y^2 + 2}{2y\sqrt{y^2 + 4}}
    \label{eq:magnification_GO}
\end{equation}

Since the GW waveform is amplified by the amplification factor given by Eq. \eqref{eq:F_SPA}, the luminosity distance of the source inferred from the waveform will be an effective distance defined by $d_{L,j}^{\rm eff}\equiv d_L/\sqrt{\mu_j}$ from each image. By taking the ratio of $d_{L,+}^{\rm eff}/d_{L,-}^{\rm eff}$, the magnification ratio between lensed images $\mu_r\equiv\mu_-/\mu_+$ can be obtained, and then the source position $y$ can be evaluated by \cite{Sereno:2011ty}
\begin{equation}
    y = \frac{1-\sqrt{\mu_r}}{1+\sqrt{\mu_r}}.
    \label{eq:y_mu_GO}
\end{equation}
In principle, once $y$ is obtained, the time delay and the magnification factor of each image can be evaluated by Eqs. \eqref{eq:time_delay_GO} and \eqref{eq:magnification_GO}, and the true luminosity distance of the source can be determined by $d_L = d_{L,j}^{\rm eff}\sqrt{\mu_j}$. 

However, with the MSD transformation in the time delay and the amplification factor given by Eqs. \eqref{eq:deltat_lambda} and \eqref{eq:F_SPA_lambda}, the effective distance is scaled by $\lambda$, but the ratio of effective distances from two images is unchanged for varying $\lambda$. As a result, the true luminosity distance cannot be determined using this approach without knowing $\lambda$. Therefore, the lens model cannot be reconstructed well from the observables of lensed images alone under the MSD in the geometric optics limit. 

On the other hand, the amplification factor for the wave-optics regime is computed exactly with
Eq. \eqref{eq:F_lambda} with different expression of $\Psi(x)$ for different lens models. 
For the PM model \cite{PhysRevD.108.043527}, the analytical solution for the amplification factor with MSD is found to be \cite{Cremonese:2021puh}
\begin{align}
    F_\lambda = & \frac{1}{\lambda}\bigg(-\frac{iw\lambda}{2} \bigg)^{1+\frac{iw\lambda }{2}} \exp \bigg[\frac{1}{2}iw \lambda^2 y^2 \bigg] \Gamma \bigg(-\frac{iw\lambda }{2} \bigg) \nonumber\\
    & \prescript{}{1}{F}_1\bigg(1-\frac{iw\lambda}{2};1;-\frac{iw\lambda}{2}y^2 \bigg),
\label{eq:F_lambda_PM}
\end{align}
where $\Gamma$ is the gamma function and $\prescript{}{1}{F}_1$ is the confluent hypergeometric function. By taking $\lambda=1$, Eq. \eqref{eq:F_lambda_PM} reduces to the form given by \cite{Nakamura:1999uwi,Matsunaga_2006}. 
Note that this expression also follows the general formula of Eq. (\ref{eq:F_lambda_sph}). 
Since the amplification factor leads to non-trivial distortions of the waveform, studying the invariance under transformations in $\lambda$ requires evaluating how the waveform changes. We will study this in detail in Sec.~\ref{sec:methods}.

The amplification factor for the PM model can also be derived in terms of Laguerre polynomials $L_n(z)$ \cite{Ezquiaga:2020spg}
\begin{equation}
    F(w,y)=\nu^{1-\nu}e^{2\nu\log\theta_E}\Gamma(\nu)L_\nu(-\nu y^2)\,,
\end{equation}
where $\nu=-iw/2$. This is equivalent to the more common representation in terms of hypergeometric functions such as in Eq.~(\ref{eq:F_lambda_PM}), except for the $e^{2\nu\log\theta_E}$ term. 
This term just redefines the reference time \cite{Ezquiaga:2020spg} and it is typically neglected.  
When we include $\lambda$, we obtain 
\begin{equation}
    F_\lambda(w,y)=\frac{e^{(1-\lambda)\nu_\lambda y^2}}{\lambda}\nu_\lambda^{1-\nu_\lambda}e^{2\nu_\lambda\log\theta_E}\Gamma(\nu_\lambda)L_{\nu_\lambda}(-\nu_\lambda y^2)\,,
\end{equation}
where $\nu_\lambda=-i\lambda w/2$. 
This expression is useful because it is then easier to generalize to the transformation from Eq.~(\ref{eq:invariance_transform}) that depends on the physical parameters: the additional surface mass density $m$, constant convergence $\kappa_c$, and the local expansion rate $h_0$. 
In terms of these parameters, and shortening as before $\lambda=m h_0$, we obtain
\begin{equation}
\begin{split}
    F_{\lambda,\kappa_c}&(w,y)=[\gamma(\lambda+\kappa_c)^{-2}]^{1+\nu_\lambda} \\
    &\cdot\frac{e^{(\gamma-\lambda)\nu_\lambda y^2}}{\lambda}\nu_\lambda^{1-\nu_\lambda}e^{2\nu_\lambda\log\theta_E}\Gamma(\nu_\lambda)L_{\nu_\lambda}(-\gamma\nu_\lambda y^2)\,,
\end{split}
\end{equation}
where $\nu_\lambda=-i\lambda w/2$ and $\gamma\equiv \lambda(\lambda+\kappa_c)^2/((\lambda+\kappa_c)^2-\kappa_c)$. 
Note that one recovers the previous expression when $\lambda=mh_0=1-\kappa_c$ (so that $\gamma=1$). 
We will use this expression when having $h_0$ as a free parameter.

\subsection{Singular Isothermal Sphere}

In addition, we also consider the Singular Isothermal Sphere (SIS) model, which represents a first order approximation to the density profile of a galaxy and its dark matter halo (see \cite{1987gady.book.....B}). This model is defined by a Newtonian potential $\Psi(x)=x$. The density profile of the SIS model is related to the velocity dispersion $\sigma_v$ by $\rho(r)=\sigma_v^2/2\pi Gr^2$. Then the Einstein radius of the SIS model is obtained by \cite{Takahashi:2016jom}
\begin{equation}
    \theta_*=4\pi \frac{\sigma_v^2}{c^2} \frac{D_{LS}}{D_S}
\end{equation}
and the redshifted lens mass in the SIS model is~\cite{Takahashi:2016jom}
\begin{equation}
    M_{L,z} = 4\pi (1+z_L)\frac{\sigma_v^4}{Gc^2} \frac{D_L D_{LS}}{D_S}.
\end{equation}

As a result, one can define the dimensionless frequency for the SIS model $w$ similarly to the PM model, which is $w=4\pi GM_{L,z}\omega/c^3$.
For a SIS model, two images form when the source position is within the Einstein radius, $\theta_S<\theta_*$. Otherwise, there is only one image. In the multiple-image region,
the time delays and the magnification factors of the images in the geometric optics limit are given by \cite{Takahashi:2016jom}
\begin{align}
& t_\pm = \mp y - \frac{1}{2}, \\
& \mu_\pm = \pm 1 + \frac{1}{y}.
\end{align}
Here, the situation is similar to the PM. From the relative magnifications, one could try to obtain $y$, which then allows fixing the arrival times and luminosity distances. However, as before, the MSD prevents from fully fixing the model parameters.

In the wave-optics limit, by inserting the potential $\Psi(x)=x$ in Eq. \eqref{eq:F_sphere}, an analytical solution (in terms of a series expansion) can be obtained as shown by Eq. (32) in \cite{Matsunaga_2006}. Under the MSD, we insert $\Psi(x)=x$ into Eq. \eqref{eq:F_lambda}. By observing that the only change in the integral is replacing $w$ by $w\lambda$, the solution of Eq. \eqref{eq:F_lambda} for the SIS model can be transformed from Eq. (32) in \cite{Matsunaga_2006} as 
\begin{align}
    F_\lambda & = \frac{1}{\lambda} \exp\bigg(\frac{1}{2}iw\lambda^2y^2 \bigg) \nonumber\\
    & \times \sum_{n=0}^{\infty} \frac{\Gamma(1+n/2)}{n!} \left(2w\lambda e^{i\frac{3\pi}{2}} \right)^{\frac{n}{2}} \prescript{}{1}{F}_1\bigg(1+\frac{n}{2};1;-\frac{iw\lambda}{2}y^2 \bigg) \,.
\label{eq:F_lambda_SIS}
\end{align}
With only one lensed GW signal, one can only reconstruct the lens model by estimating the parameters from the lensed waveform itself. In the next section, we will investigate the method to break the MSD with waveform analysis in the wave-optics regime.

\section{Inferring the GW and lens properties}
\label{sec:methods}

After presenting the general impact of the invariance transformations on the GW observables of lensed signals, we now explore how they affect the inference of GW and lens properties in practice. 
We perform studies with different data analysis techniques of increasing level of complexity, starting with a template mismatch analysis, followed by a Fisher information matrix calculation, and concluded with full parameter estimation via Bayesian inference.

\subsection{Mismatch}

The match between two GW signals is defined by maximizing their overlap over time $t$ and phase $\psi$:
\begin{equation}
    {\rm match} \equiv {\rm max} \bigg| \frac{(h_1|h_2)}{\sqrt{(h_1|h_1)(h_2|h_2)}}\bigg|_{t,\psi}\,,
\end{equation}
where the noise-weighted inner product of two real, time-domain signals is defined by
\begin{equation}
  (h_1|h_2) \equiv 2\int_{0}^{\infty} \frac{h_1(f)h_2^{*}(f)+h_1^{*}(f)h_2(f)}{S_n(f)} {\rm d}f\,.
\end{equation}
Thanks to the normalization, if two signals are identical, the match is 1. On the contrary, the mismatch is defined as $\epsilon=1-$match. 
The mismatch is an interesting statistic for two main reasons. It measures the differences between two signals independently of their amplitude, or SNR $\rho=\sqrt{(h|h)}$. Moreover, in the limit of a high-SNR signal in some data, the mismatch $\epsilon$ can be related to the improvement in the $\chi^2$ goodness-of-fit to the data of one model versus the other, $\Delta\chi^2\approx 2\epsilon\rho^2$ (see e.g. \cite{Ezquiaga:2022nak}). Therefore, a high mismatch indicates that it would be possible to clearly prefer one hypothesis over the other.

\begin{figure}
    \centering
    \begin{tabular}{@{}c@{}}
    \includegraphics[width=0.48\textwidth]{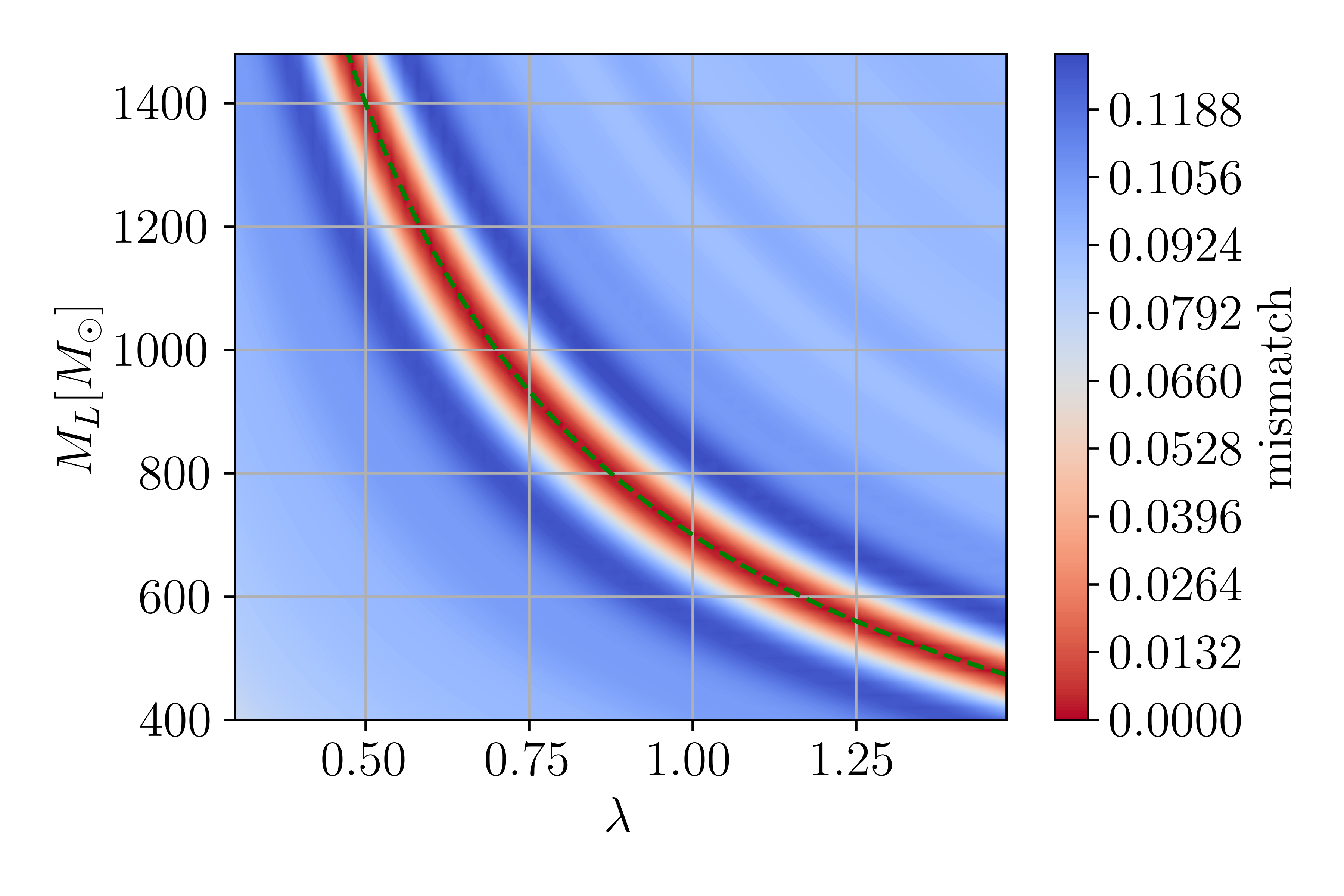}
    \end{tabular}
    \begin{tabular}{@{}c@{}}
    \includegraphics[width=0.48\textwidth]{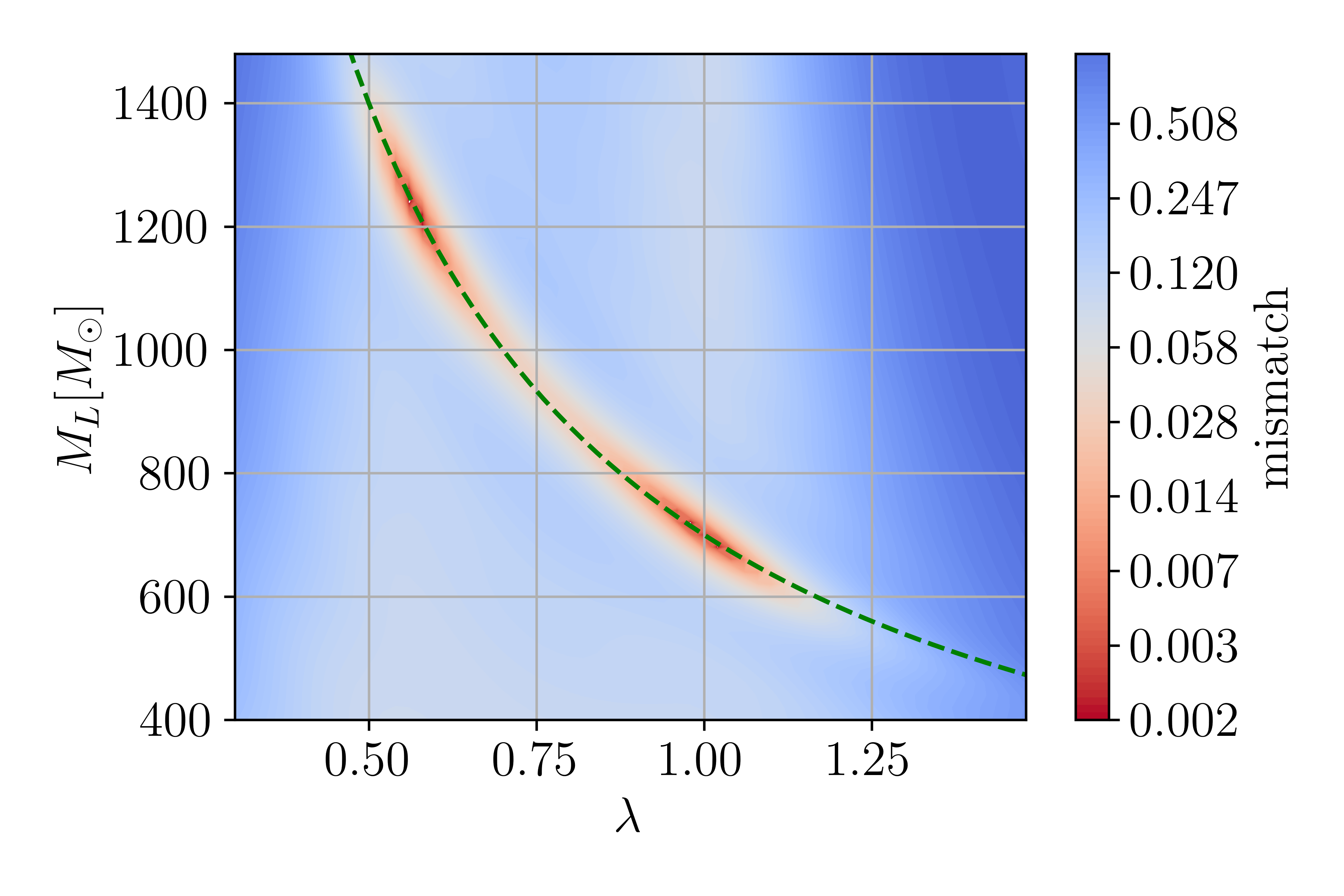}
    \end{tabular}
    \caption{Mismatch between time-aligned lensed waveforms for different $M_L$ and $\lambda$ and the injected lensed waveform for $M_L=700\,M_\odot$ and $\lambda=1$. The injected parameters for the source are listed in Table \ref{tab:injpar} (left column). The green dashed line indicates $M_L=M_{L,\mathrm{fid}}/\lambda$. Upper panel: The match is aligned with time shifts computed with Eq. \eqref{eq:t_lambda} at each ($M_L$, $\lambda$) point. Lower panel: The match is aligned with a fixed time shift computed with the fiducial $M_L$ and $\lambda$.}
    \label{fig:mismatch_ML_lambda}
\end{figure}

Because lensed waveforms have an arrival time delay, they need to be aligned in time first to maximize the match between them. For the lensed waveforms we generate at different points in the ($M_L$, $\lambda$) space while fixing other parameters, the only time differences between each waveform is this arrival time delay due to lensing, so the maximization of matches over time shift can be achieved by removing this time delay computed at each ($M_L$, $\lambda$) point.
To do so, the amplification factor needs to be multiplied by a time shift term $\exp (-i w t_{\lambda})$, which is expanded using Eq. \eqref{eq:t_lambda} in dimensionless unit as
\begin{align}
    \exp(-iw t_\lambda) & = \exp\bigg[-iw\bigg(\lambda t -\frac{\lambda(1-\lambda)}{2}y^2 \bigg) \bigg] \nonumber\\ 
    & = \exp\bigg[-iw\lambda\bigg(t-\frac{y^2}{2} \bigg) \bigg]\exp\bigg(-\frac{1}{2}iw\lambda^2y^2 \bigg) \,.
\label{eq:time_align}
\end{align}
The term $\exp(-iw\lambda^2y^2/2)$ in the time alignment cancels the term $\exp(iw\lambda^2y^2/2)$ in the analytical solution of $F_\lambda$ in Eq. \eqref{eq:F_lambda_PM}, so that only the terms with $w\lambda$ are left in $F_\lambda$, which makes it impossible to distinguish $M_L$ and $\lambda$ through comparing GW waveforms since \mbox{$w=8\pi(1+z_L)M_Lf$}.
Therefore if one were to align the waveforms with parameter-dependent time delay in Eq. \eqref{eq:time_align} with exact values of $M_L$ and $\lambda$, the MSD could not be broken for the PM model. We illustrate the degeneracy in the upper panel of Figure \ref{fig:mismatch_ML_lambda}, where the mismatch is computed using the LIGO A+ design sensitivity curve \cite{aLIGO_design_curve}. 
There is a band of zero mismatch that tracks the expected $M_L=M_{L,\mathrm{fid}}/\lambda$ scaling over the whole range of $\lambda$, where the fiducial values are $M_{L,\mathrm{fid}}=700M_\odot$ and $\lambda_{\mathrm{fid}}=1$.

However, in the actual observations, the exact time delay of an image cannot be obtained by using Eq. \eqref{eq:F_lambda_PM}, since the lens parameters $M_L$, $y$ or $\lambda$ are not known. 
The only observable from the GW signals is the coalescence time $t_c$. 
Therefore, when performing parameter estimation (which we will show in section \ref{subsec:PE}), the waveform cannot be shifted by a time delay computed by values of ($M_L$, $\lambda$) at different parameter-space points.
Instead, we compute the mismatch between the signals shifted by the same amount of time for all parameter-space points to illustrate the scenario where the exact time shift is unknown.
We choose this fiducial time shift to be the value of the time delay of the fiducial waveform. Then we shift all waveforms constructed with various ($M_L$, $\lambda$) with this fiducial time shift.
As shown in the lower panel of Figure \ref{fig:mismatch_ML_lambda}, the MSD is now broken in the mismatch plot. 
The region with low mismatch still follows $M_L\propto 1/\lambda$ but it is now bounded.
Although theoretically one can search for a value of $t_c$ that maximizes the matches over time shift, it now also depends on parameters other than ($M_L$, $\lambda$). 
This shows that the correlation between the arrival time and the waveform distortion is key in breaking this degeneracy. As we will see in Section \ref{subsec:PE}, in parameter estimation for detected signals in a multi-parameter space, the degeneracy appears between $t_c$ and lensing parameters, but the degeneracy between $M_L$ and $\lambda$ becomes bounded.

\subsection{Fisher Matrix}

For the next analysis, we assume that the noise is stationary and Gaussian and that we are in the limit of high SNR, so that the likelihood peak becomes narrow enough that the prior around the maximum-likelihood value $\theta_{ML}^i$ can be taken as approximately flat.
Then the posterior distribution of the parameters $\theta^i=\theta_{ML}^i+\Delta\theta^i$ given the data $d$ is obtained by \cite{PhysRevD.46.5236, PhysRevD.49.2658}
\begin{equation}
    p(\theta^i|d) = {\cal N} e^{-\frac{1}{2}\Gamma_{ij}\Delta \theta^i \Delta \theta^j},
\end{equation}
where $\Gamma_{ij}$ is the Fisher information matrix defined by
\begin{equation}
    \Gamma_{ij} \equiv \bigg( \frac{\partial h}{\partial \theta^i} \bigg| \frac{\partial h}{\partial \theta^j} \bigg),
\end{equation}
and ${\cal N}=\sqrt{{\rm det(\Gamma/2\pi)}}$ is the normalization factor. 
In this approximation, the deviations from the maximum likelihood $\Delta\theta^i$ are small.
The uncertainties in the parameters can be obtained from the covariance matrix, which is just the inverse Fisher matrix. Then, 
the variance of the parameter $i$ is
\begin{equation}
    \sigma_i^2= \left(\Gamma^{-1}\right)^{ii}~.
\end{equation}
Therefore, by computing the derivatives of the lensed signal with respect to each parameter numerically, we compute the Fisher matrix to forecast the constraints of the parameters from lensed GW signals.

\begin{figure}
    \centering
    \includegraphics[width=0.5\textwidth]{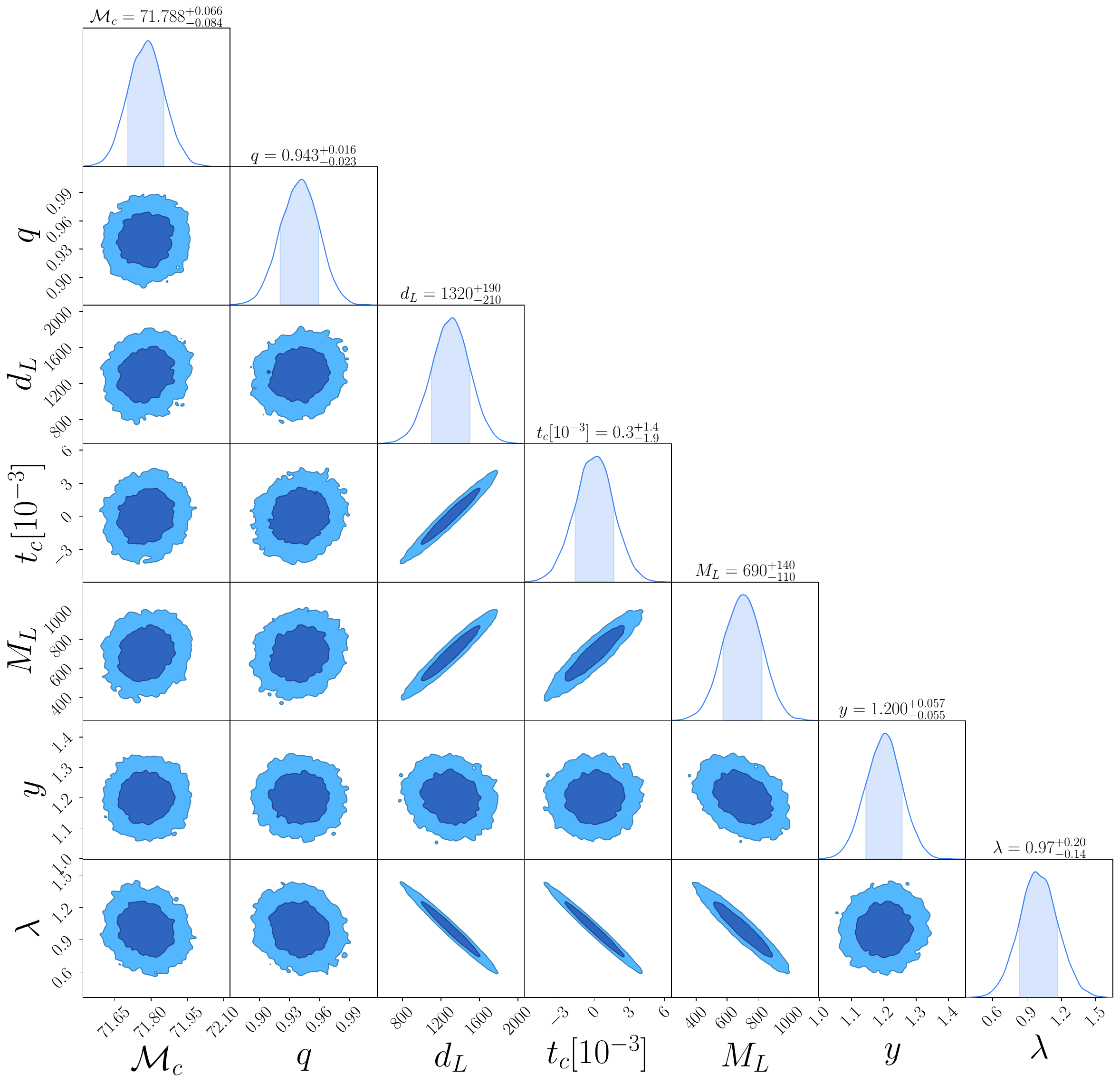}
    \caption{Fisher matrix forecast for the lensed waveform in the wave-optics regime without time alignment. The injected parameters for the source are listed in Table \ref{tab:injpar}. }
    \label{fig:fisher_noalign}
\end{figure}

We plot the contours for the Fisher matrix constraints on the parameter set $\{{\cal M}_c, q, d_L, t_c, M_L, y, \lambda \}$ in Figure \ref{fig:fisher_noalign}, where ${\cal M}_c=(m_1m_2)^{3/5}/(m_1+m_2)^{1/5}$ is the chirp mass; $q=m_2/m_1$ is the mass ratio of the binary source with component masses $m_1$ and $m_2$ ($m_2<m_1$); $d_L$ is the luminosity distance of the source; $M_L$ is the mass of the lens; and $y$ is the impact parameter. Here, and in the whole paper, we refer to the detector-frame parameters, so, for example, $M_L$ stands for the mass of the lens with redshift applied from the detector frame.
The injected values for the binary source parameters are listed in Table \ref{tab:injpar} (left column), which is the same as in Figure \ref{fig:mismatch_ML_lambda}. 
We used the LIGO A+ design sensitivity curve \cite{aLIGO_design_curve} to compute the Fisher matrix.  
As shown in Figure \ref{fig:fisher_noalign}, the contours of the Fisher matrix forecast for several pairs of parameters are very elongated, indicating strong degeneracy between each pair of them. 
The ellipses are however finite, indicating as well that the degeneracy is broken at some point.
The MSD is shown by the contour for $(M_L,\lambda)$, which matches the results shown by the mismatch analysis in Figure \ref{fig:mismatch_ML_lambda}. Besides, the luminosity distance $d_L$ and the coalescence time $t_c$ are both degenerate with $M_L$ and $\lambda$. This is because the amplification factor is proportional to $1/\lambda$ as shown in Eq. \eqref{eq:F_lambda_PM}, and the waveform is proportional to $1/d_L$, and thus $d_L$ is degenerate with $\lambda$. And because of the degeneracy between $M_L$ and $\lambda$, $d_L$ is also degenerate with $M_L$. On the other hand, $t_c$ shifts the waveform in time by multiplying the waveform by $\exp(2\pi f t_c)$. However, the term $\exp(i\omega \lambda^2 y^2/2)$ in the amplification factor in Eq. \eqref{eq:F_lambda_PM} gives an equivalent effect to a time shift, so it creates degeneracy between $t_c$ and $M_L$, and $t_c$ and $\lambda$.

\subsection{Bayesian Parameter Estimation}
\label{subsec:PE}

Now, we focus on investigating the MSD using Bayesian parameter estimation (PE) analysis. In doing so, on one hand, we get a more realistic view of the behavior of the degeneracy and how it is spread across all the parameters involved. 
On the other hand, we can validate the lower-cost studies done through the Fisher matrix. 
The Fisher matrix corresponds to PE in the limit of high SNRs, taking the leading order Gaussian behavior. 
Therefore, in general, the Fisher matrix is incapable of reproducing highly non-Gaussian posterior distributions.

We run PE analyses considering both the case with a fixed $\lambda=1$ and with $\lambda$ as a free parameter. 
Apart from understanding how the degeneracy works and how it is spread across parameters, this gives precious information on what we are missing when not considering the MSD in lensed PE analyses. 

We present three main cases: 
(i) an injected event with high SNR and a PM lens model; 
(ii) an injected event with high SNR and an SIS lens model; and 
(iii) an event similar to the real GW200208\_130117~\cite{KAGRA:2021vkt} (from here on GW200208 for short). 
For the first two cases, we injected a regular and a biased signal, i.e. with \mbox{$\lambda=1$} and \mbox{$\lambda\neq1$}, respectively. The details of the parameters used in these cases can be found in Table~\ref{tab:injpar}, and more information about the software used can be found in Appendix~\ref{ap:PE}.

{\renewcommand{\tabcolsep}{1.5mm}
{\renewcommand{\arraystretch}{1.5}
\begin{table}[th]
    \centering
    \caption{Injection parameters for the cases with the PM and SIS lens models and for the GW200208-like event. 
    }
    \begin{tabular}{|c|c|c|c|}
    \hline
    \textbf{parameter} & \textbf{PM} & \textbf{SIS} & \textbf{GW200208} \\
    \hline \hline
        $\mathcal{M}_c$ [$M_\odot$] & $71.78$     & $51.36$&  $38.97$ \\    \hline
        $q$ &                       $0.94$      & $0.71$ &  $0.77$  \\    \hline
        $d_L$ [Mpc] &               $1300$      & $2500$ &  $2294$  \\    \hline
        $\cos\theta_{JN}$ &         $0.95$      & $0.98$ &  $-0.83$ \\    \hline
        $M_L$ [$M_\odot$] &     $700$       & $1500$ &  $1693$  \\    \hline
        $y$ &                       $1.2$       & $0.9$  &  $1.47$  \\    \hline
        $\lambda$ &                 $1/0.8$     &$1/0.85$&  $1$     \\    \hline
        detectors &                 \multicolumn{3}{c|}{H1, L1, V1}  \\ \hline
        optimal SNR &               $78$        & $35.45$&  $19.12$ \\    \hline
        waveform    &               \multicolumn{3}{c|}{\multirow{2}{*}{IMRPhenomXPHM~\cite{Pratten:2020ceb}}} \\ 
        approximant & \multicolumn{3}{c|}{}                      \\ \hline
    \end{tabular}
    \label{tab:injpar}
\end{table}}}

For each of these three cases, we did (at least) two PE analyses, one with fixed $\lambda$ and one with it free. 
Moreover, to better understand the propagation of the degeneracy, and depending on the situation/event, we also took into account different sub-cases where we fixed the value of other parameters. 
For clarity, we present and discuss here the three main corner plots that we get from these analyses. Other valuable plots and discussions can be found in Appendix~\ref{ap:PE}, along with more details about the injections.

\subsubsection{Point mass lens model}
\label{subsec:PM_model}

\begin{figure}[htbp]
    \centering
    \includegraphics[width=.49\textwidth]{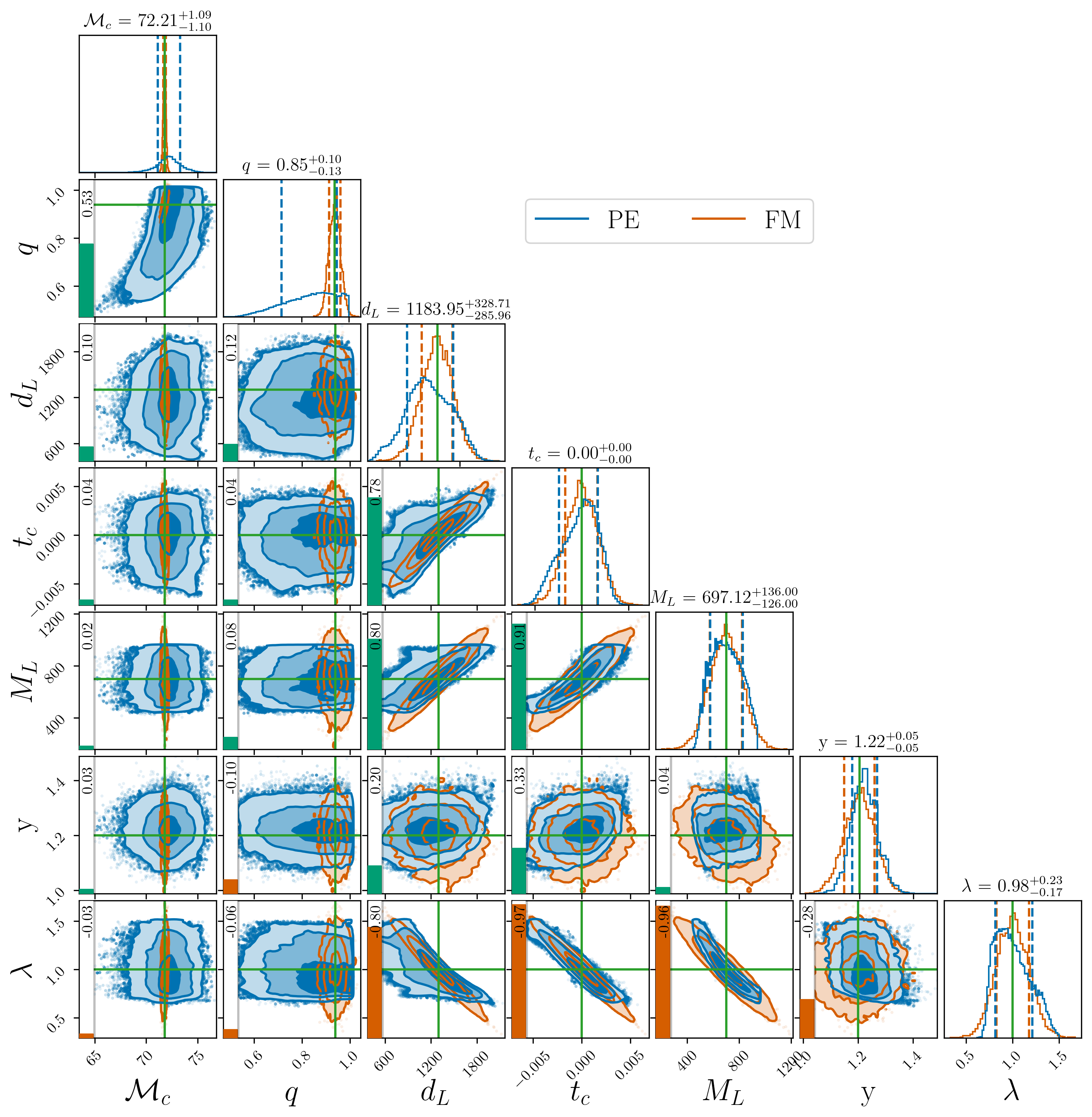}
    \caption{
        Posterior distributions from Bayesian PE (blue) and Fisher matrix (FM, red) analyses for the case with an injected GW signal under the PM lens model and with $\lambda=1$, and recovered with $\lambda$ as a free parameter.
        The bars, and the numbers inside them, on each sub-panel indicate the covariance between the parameters. When green, the covariance is positive, and negative when red.
    }
    \label{fig:PEvsFM_PM}
\end{figure}

The first case we want to analyze, i.e. the PM lens model, is the most straightforward and the most suitable for the range of lens masses that we are considering. In \figurename~\ref{fig:PEvsFM_PM}, we present the main result for this case. In particular, we show the case of the injection with a PM lens and $\lambda=1$ (see Tab.~\ref{tab:injpar}), recovered through a PE analysis where we left $\lambda$ as a free parameter.
Ignoring the orange contours for the moment, we can see in blue the posterior distribution of the main parameters of the event. 

The main aspect to notice is the absence of correlation between the lens mass, $M_L$, and the impact parameter, $y$. These parameters should correlate under the MSD, giving place to the degeneracy. They do so in the ``regular'' case when $\lambda$ is fixed to $1$ in the PE analysis, i.e. when we do not take the MSD into consideration. The absence of correlation here indicates that we are tracing the MSD correctly. But it is interesting to notice where this correlation has shifted to. On the one hand, as we expect, we see a strong correlation between $M_L$ and $\lambda$; on the other hand, though, the correlation of $\lambda$ with $y$ is not as high, and we can see that the luminosity distance, $d_L$, strongly correlates with it. This can be easily explained by looking at Equation~\eqref{eq:F_lambda_sph}, where we see that the amplitude of the amplification factor scales like $1/\lambda$, a behavior that can be easily mimicked by changing the luminosity distance inversely to $\lambda$.

\subsubsection{SIS lens model}
\label{subsec:SIS_model}

\begin{figure}[htbp]
    \centering
    \includegraphics[width=.49\textwidth]{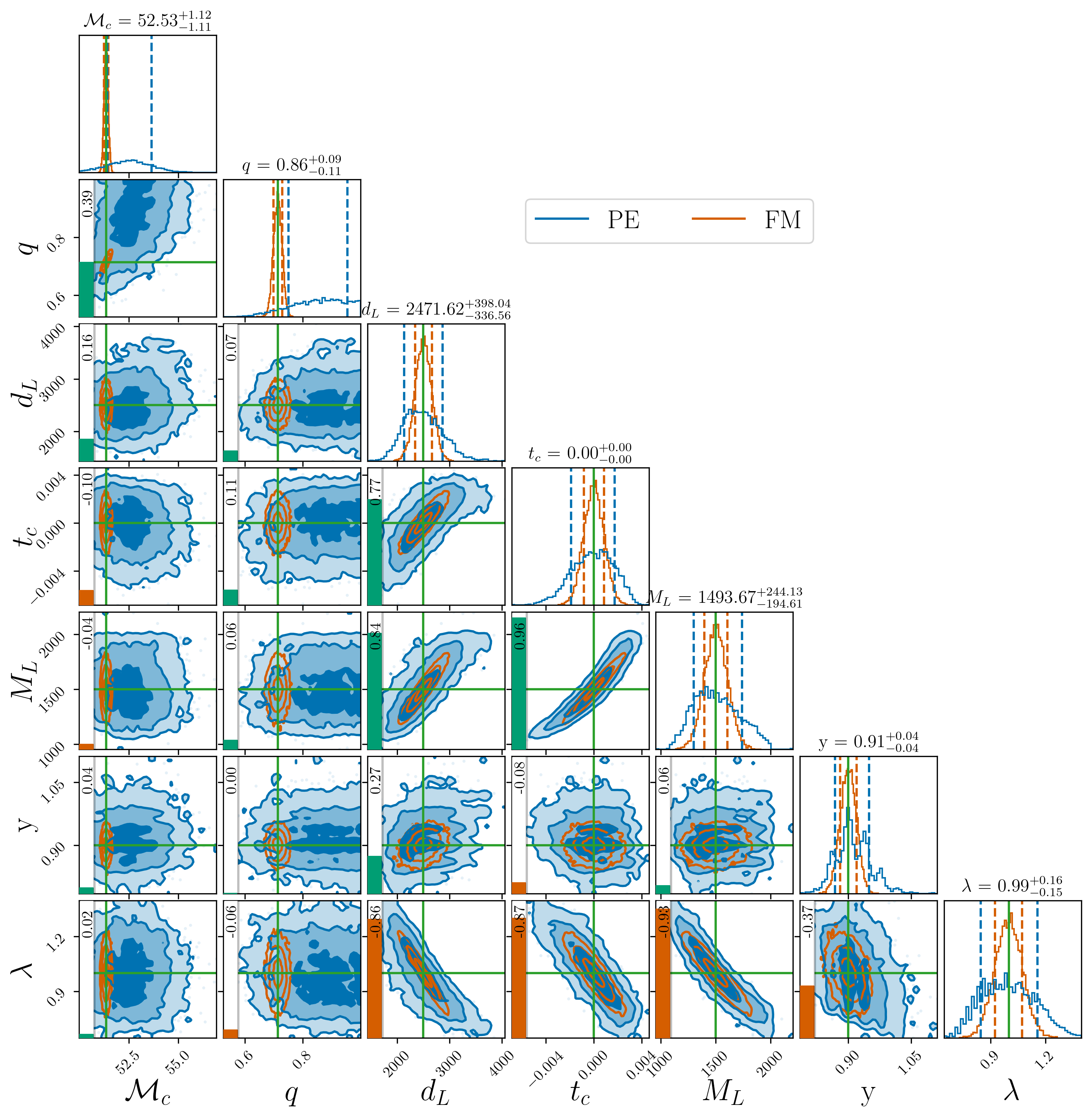}
    \caption{
        Posterior distributions from Bayesian PE (blue) and Fisher matrix (FM, red) analyses for the case with an injected GW signal under the SIS lens model and with $\lambda=1$, and recovered with $\lambda$ as a free parameter.
    }
    \label{fig:PEvFM_sis}
\end{figure}

Next, we consider a (slightly) more complex lens model, the SIS. The injection parameters are the ones in Tab.~\ref{tab:injpar}, and the PE analysis has again been done leaving $\lambda$ free.
From \figurename~\ref{fig:PEvFM_sis}, just like for the PM case, we can see that the expected $M_L-y$ correlation has been transferred to one between $M_L$ and luminosity distance, as they both strongly correlate with $\lambda$.

An interesting aspect affecting both cases, PM and SIS, is that, while all other parameters are well constrained around their injected value, we notice that those that correlate the most with $\lambda$ have rather high uncertainties\footnote{Especially when compared to the case where we fix $\lambda=1$ during the PE analysis. See appendix~\ref{ap:PE} for more details.}. This is due to the fact that we add a new parameter to the analysis and its implication will be discussed in Section~\ref{sec:implications_astro}.

\subsubsection{GW200208-like event}
\label{subsec:GW200208}

While both previous cases have high SNRs, we present now a more typical scenario for current ground-based detectors. This analysis is done on an injection that follows the characteristics of the real event GW200208\footnote{N.B.: The real event is not considered to be lensed, but it raised interest in the community as a possible lensed candidate. We use here the parameters that give the most similar waveform to GW200208 if it were lensed.}, that are listed in Tab.~\ref{tab:injpar}. As the lens mass is still rather small, we decide to use the PM lens model both for injection and PE analysis.
This event gives us the possibility to see how the PE and MSD behave in a scenario that is more realistic given the current detected events.

\begin{figure}[thbp]
    \centering
    \includegraphics[width=.49\textwidth]{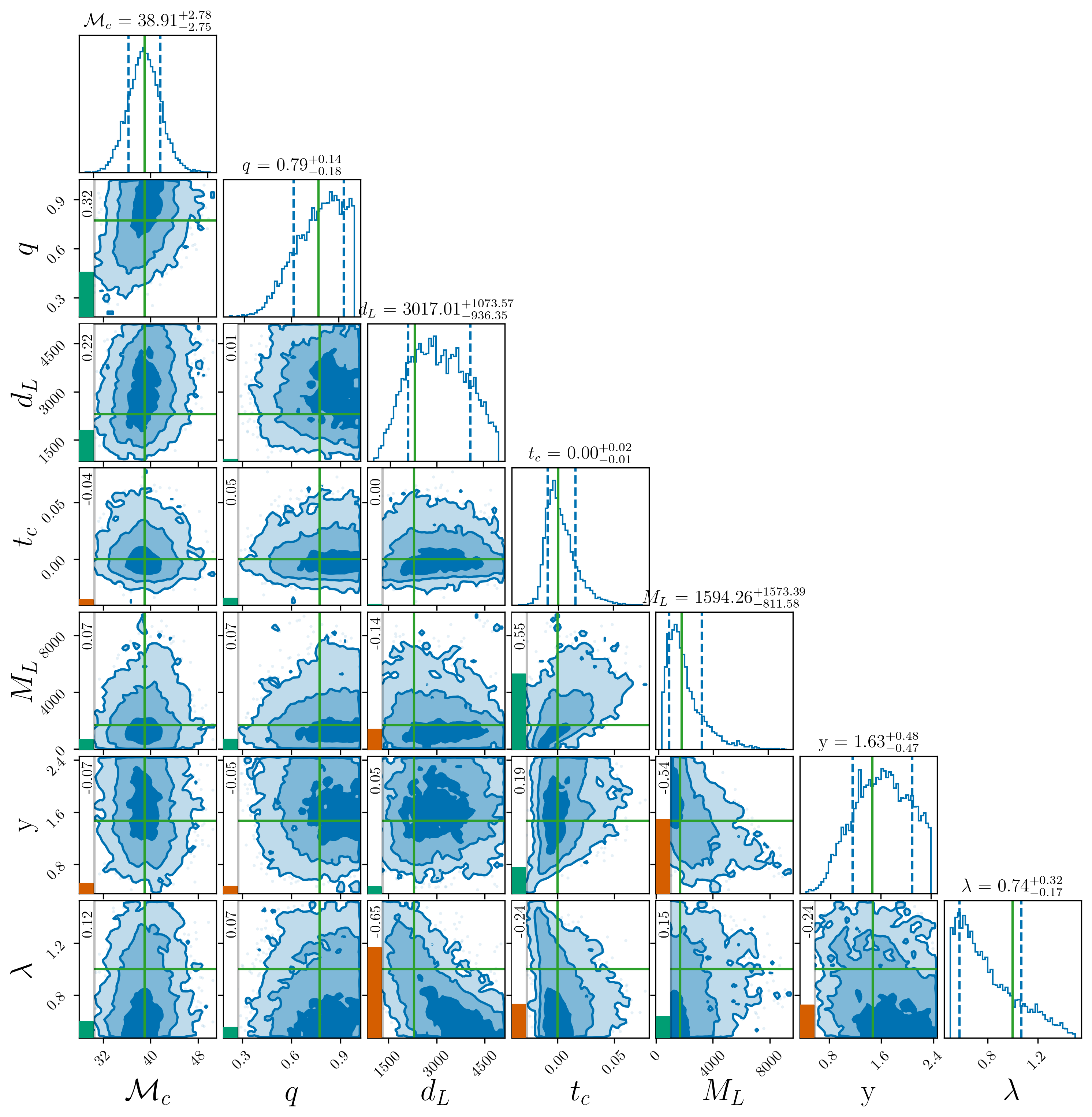}
    \caption{
        Posterior distributions from Bayesian PE analysis for the case with a GW200208-like injection (with a PM lens model) and recovered with $\lambda$ as a free parameter.
    }
    \label{fig:gw200208-lam}
\end{figure}

Contrary to the previous cases, we can immediately see that all parameters are constrained more poorly. This is due to the lower SNR of the event. It follows that also the correlations between the parameters do not adhere to our expectations from the previous cases. In particular, the correlation $M_L-y$, between the lens parameters, is still rather high and has not been transferred to $M_L-\lambda$ and $d_L-\lambda$. Overall, in this case it is difficult to draw interesting conclusions, except that it seems unrealistic to expect any decisive constraints on lens parameters (and consequently on cosmology) from this kind of events. We will expand on this aspect in Section~\ref{sec:implications_cosmo}.

\subsubsection{Fisher Matrix validation}
\label{subsec:FM_val}

Finally, we want to compare the study made through the Fisher matrix (FM) with the more complete PE analysis. We do so to validate the FM results and exploit its convenience in the following calculations to study a larger number of cases than we could using only the PE method.

In \figurename~\ref{fig:PEvsFM_PM} and \ref{fig:PEvFM_sis}, we compare the results of these two methods for the injected events described in Tab.~\ref{tab:injpar}.
In blue, we see the posterior distributions from the PE analysis, and in orange those from the FM. 
The main differences are found in the chirp-mass and mass-ratio posteriors. Here, the FM results are much narrower but, since we are interested in the lens parameters, this does not change anything for our study.
Another difference can be seen comparing \figurename~\ref{fig:mismatch_ML_lambda} and \ref{fig:fisher_noalign} where we see that the shape of the $M_{L}-\lambda$ degeneracy (that goes like $M_{L}\propto1/\lambda$) is tracked well by the mismatch analysis, but not as much by the FM. This is due to the nature of the FM, which takes the (value of the) derivative of the model (in this case the amplification factor) in a single point. Nonetheless, in \figurename~\ref{fig:PEvsFM_PM} and \ref{fig:PEvFM_sis} the degeneracy is tracked well and, apart from these differences, by the high overlap between the two posterior distributions for all parameters, for both PM and SIS cases, we can safely say that, in this wave optics regime, the FM can be used instead of the more resource-consuming PE.

\section{Implications for astrophysics}
\label{sec:implications_astro}

In this section, we study the implications of the MSD for astrophysical applications of observing lensed GWs. The two main effects that we are going to discuss are 
(i) increased uncertainties in lens reconstruction; and
(ii) biases on the lens model inference.
As we shall see, these aspects are tightly intertwined and will be presented together, highlighting the common causes. 

\begin{figure}[htbp]
    \centering
    \includegraphics[width=.49\textwidth]{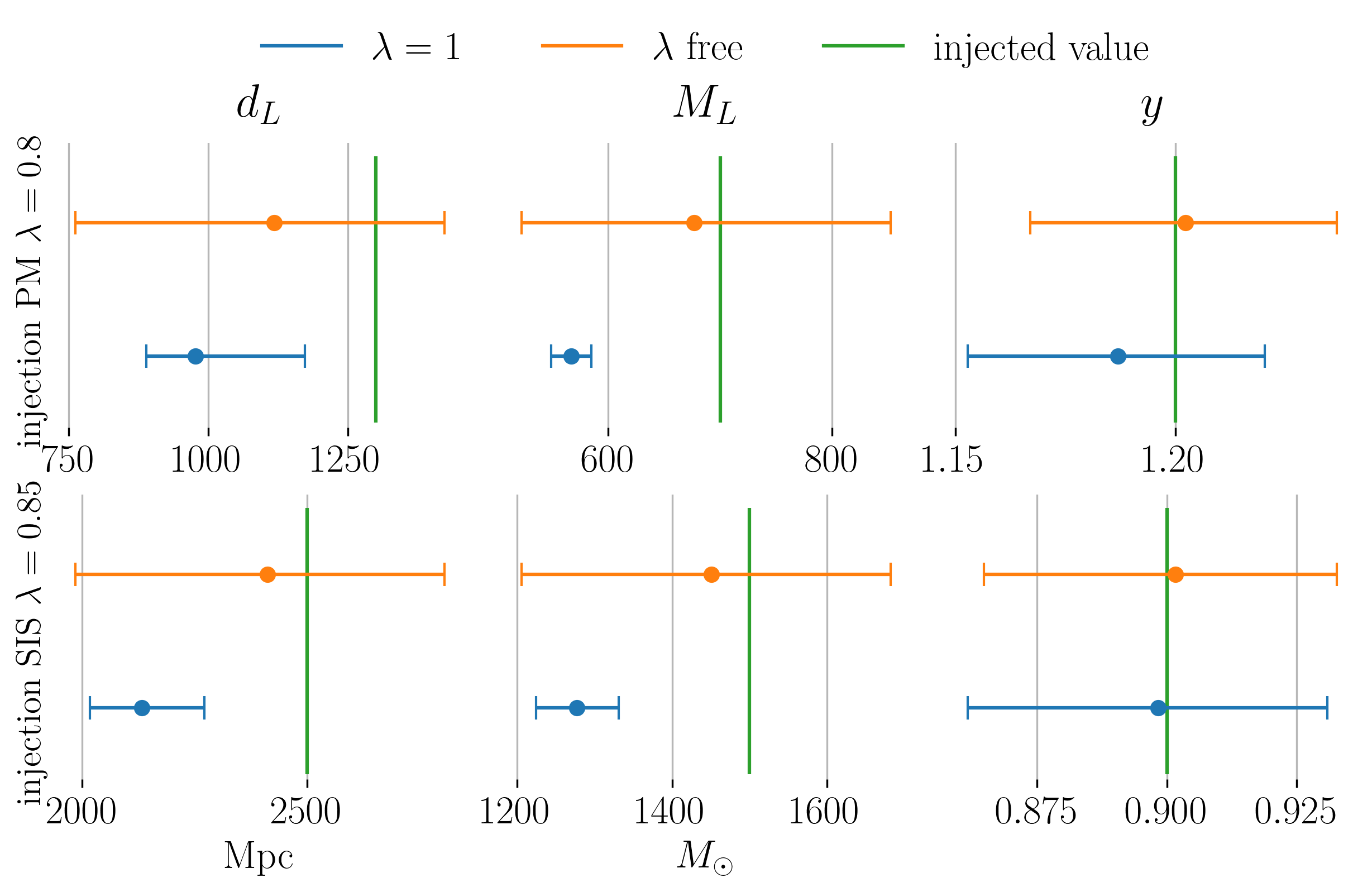}
    \caption{
        Values of luminosity distance, redshifted lens mass, and dimensionless impact parameter for the cases of a point mass (PM) and singular isothermal sphere (SIS) with injected $\lambda\neq1$. 
        In blue the PE with $\lambda=1$, in orange that with $\lambda$ free.
    }
    \label{fig:values_recap}
\end{figure}

The first and main problem that the MSD brings into the game is the biases on (some of) the parameters we want to retrieve from the signal. 
We are summarizing this problem in \figurename~\ref{fig:values_recap}\footnote{
    See also \Cref{fig:lam08-nolam,fig:lam08-lam,fig:sislam085-nolam,fig:sislam085-lam} in Appendix~\ref{ap:PE} for a longer discussion.
}. 
Here, for two injections for PM (top row) and SIS (bottom row) lenses with $\lambda\neq1$ and the three most interesting parameters (luminosity distance, $d_L$; redshifted lens mass, $M_L$; and dimensionless impact parameter, $y$, respectively, for the three columns) we show the injected values as vertical green bars; the values retrieved from a PE analysis not considering the MSD (i.e. fixing $\lambda=1$) as horizontal blue bars; and the values from the PE analysis considering the MSD (i.e. leaving $\lambda$ as a free parameter) as horizontal orange bars.

First of all, we can see that the uncertainties associated with the first two parameters increase when we consider the MSD, i.e. in the orange bars. This is not the case, though, for the impact parameter, as we saw that it does not correlate much with $\lambda$. These are well-expected behaviors and will be discussed further below.

Most importantly, we see that the parameters retrieved without considering the degeneracy are significantly biased w.r.t. the injected ones. 
This is particularly true for the luminosity distance and, even more, the lens mass. 
Again, since the value of $y$ does not correlate much with $\lambda$, also its value does not change and is not biased by the degeneracy.

\begin{figure}[htbp]
    \centering
    \includegraphics[width=.49\textwidth]{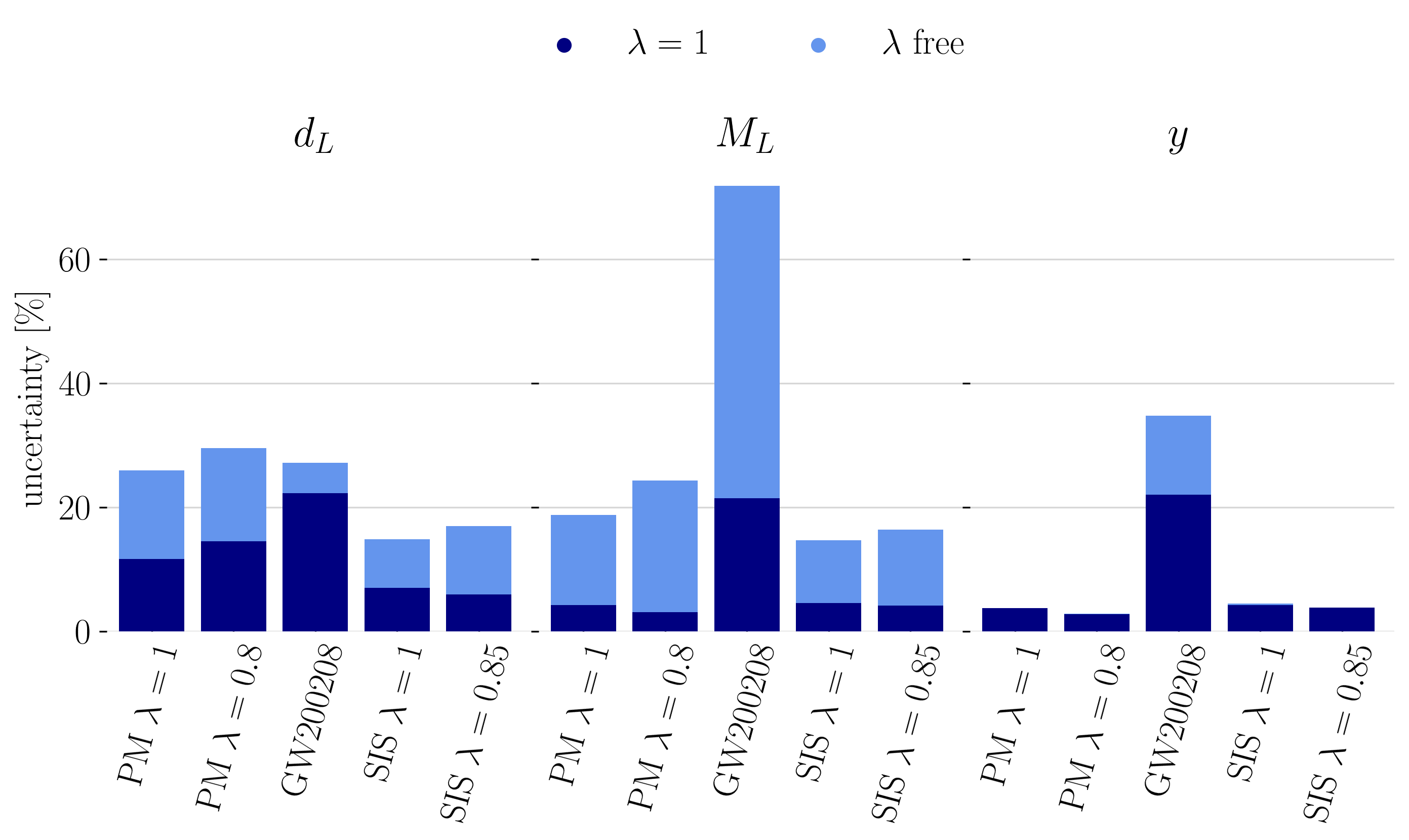}
    \caption{
        Comparison of PE uncertainties (as a percentage of the value) on the luminosity distance, lens mass, and impact parameter for the different cases of fixing $\lambda=1$ (dark blue) and leaving $\lambda$ as a free parameter (light blue).
    }
    \label{fig:errors_recap}
\end{figure}

Given the MSD, this behavior is expected since the degeneracy introduces biases in the measurement of the parameters associated with it and might lead to misleading results if not properly accounted for. This, then, will translate into an increased difficulty in selecting the most appropriate lens model.
In \figurename~\ref{fig:errors_recap} we summarize, from several injection cases, the PE uncertainties (as a percentage of the value) on the parameters of highest interest regarding the degeneracy, i.e. the luminosity distance, $d_L$, the mass of the lens, $M_L$, and the impact parameter, $y$. For each, we again compare PE neglecting the MSD, i.e. fixing $\lambda=1$, with runs accounting for the MSD, i.e. with $\lambda$ as a free parameter.

We can immediately see that leaving $\lambda$ free yields higher uncertainties. This summarizes very well the point of this study: not considering the MSD will bring lower -- but unrealistic -- uncertainties since we should always consider the presence of the degeneracy. Doing so will necessarily increase the uncertainties, but will give us a more realistic estimate of the parameters with less bias.

\begin{figure}[htbp]
    \centering
    \includegraphics[width=.49\textwidth]{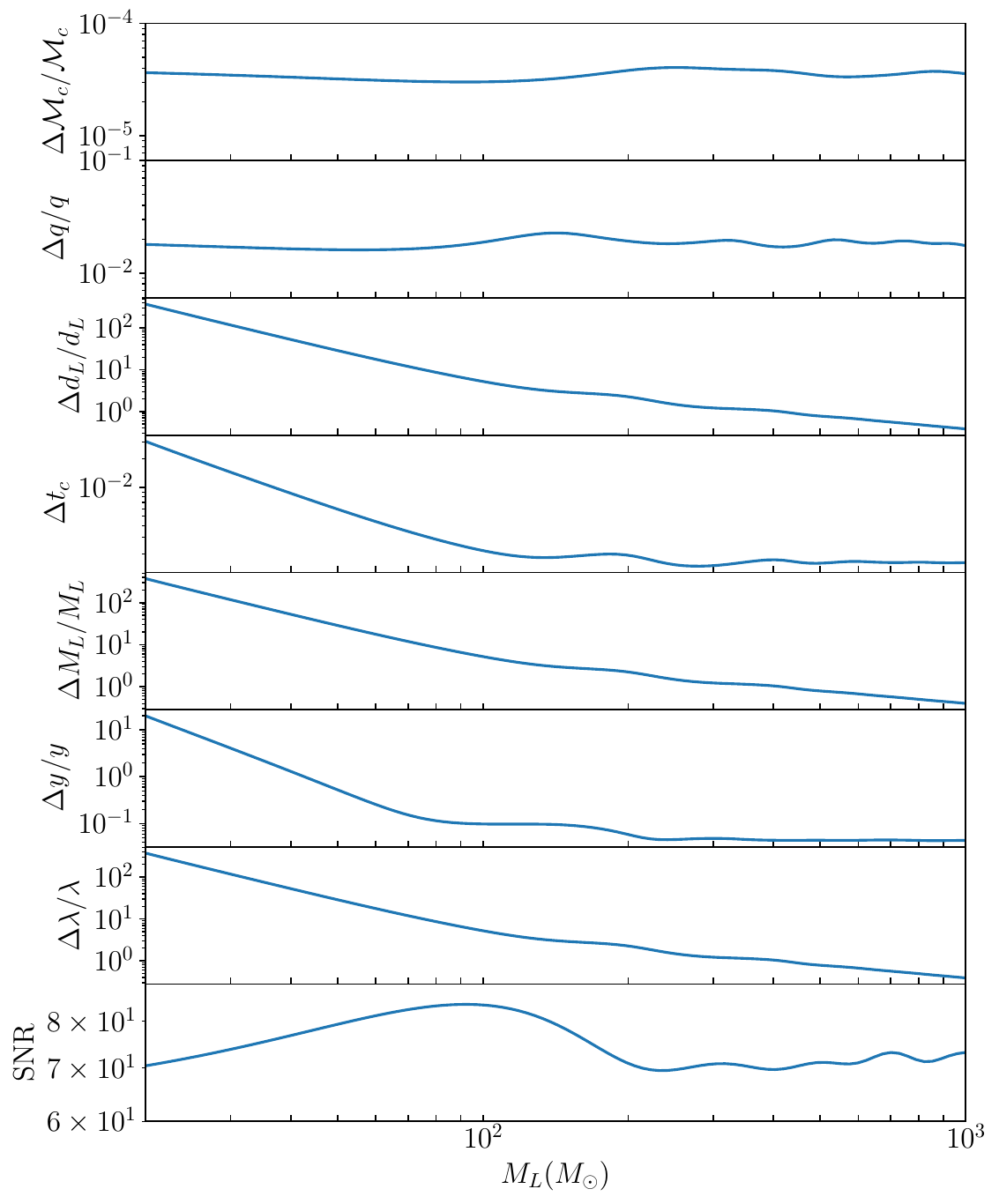}
    \caption{Relative uncertainties of parameters from a Fisher matrix forecast for different lens masses with the PM model.
    }
    \label{fig:Fisher_error_vary}
\end{figure}

Although the cases considered with full PE show a clear tendency to increase uncertainties when including the MSD, they only represent a handful of examples. 
For this reason, we use the Fisher matrix approach to scan a larger part of the parameter space in \figurename~\ref{fig:Fisher_error_vary}. 
We find that for smaller lens masses, the relative uncertainties of parameters become larger. 
For a lower frequency boundary around $f=20$ Hz for LVK detectors,
the condition on the lens mass for the geometric optics approximation to break down is $M_L<10^5\,M_\odot [(1+z_L)f/{\rm Hz}]^{-1} \approx 5000\,M_\odot$. However, when $M_L$ approaches this limit, the MSD covers an extremely elongated area in the $M_L-\lambda$ parameter space along the constant $M_L\lambda$ curve, as in the geometric optics regime. As a result, the uncertainties forecast by the Fisher matrix are not able to reveal the exact degeneracy. Therefore we choose to analyze the changes in relative uncertainties with the Fisher matrix while varying the lens mass up to $10^3\,M_\odot$ as a safe limit. As we can see, the relative uncertainties of the chirp mass and the mass ratio remain steady when increasing $M_L$, but the relative uncertainties of other parameters decrease for increased $M_L$. This indicates that uncertainties in Bayesian PE will be smaller for events with stronger lensing effects by heavier lenses in the wave-optics regime.

Finally, an important point to notice, looking at \figurename~\ref{fig:Fisher_error_vary}, is the case of very low-mass lenses, i.e. the left part of the plot. Here, we are in a situation where there is basically no lensing, as the GW wavelength is much larger than the lens itself and there are only weak lensing effects.
Therefore, we would only expect the luminosity distance to be slightly affected by lensing. 
However, we are not seeing that in \figurename~\ref{fig:Fisher_error_vary}. 
This is because the MSD parameter $\lambda$ is degenerate with the luminosity distance and there are large correlations that inflate the uncertainties.
Small lenses have little directly measurable influence on the waveform phase evolution (hence the expected large uncertainties on $M_L$ and $y$), but there could be a strong mass sheet (with $\lambda\ll 1$) that creates a large magnification [see Eq.~\eqref{eq:F_SPA_lambda}] so that the signal appears nearby. 
This would be particularly troublesome because it would bias the estimation of the source distance in this configuration without a simple way to solve the problem.

\section{Implications for cosmology}
\label{sec:implications_cosmo}

We now explore the implications for potential applications related to measuring the expansion rate of the Universe.

\subsection{Invariance transformations and $H_0$ in wave-optics lensing}
\label{sec:invariance_transform_H0}

In Sec.~\ref{sec:multiple_image}, we have presented the invariance transformation in the deflection angle in the regime of strong lensing when multiple images form. 
As shown in Eq. \eqref{eq:invariance_transform}, the scaled Hubble constant $h_0=H_0/H_0^{\rm fid}$ and the surface mass density scaling factor $m$ can both transform $\vec\alpha$ in a linear pattern. 
Therefore if we try to obtain $h_0$ from the time delay between the multiple images it will be affected by the MSD in gravitational lensing. 
For GWs, in addition, since the luminosity distance is proportional to $H_0^{-1}$, a change in $h_0$ will cause $d_L \to d_L/h_0$. So $h_0$
affects the waveform amplitude and the lensing amplification factor at the same time, which leads to a strong degeneracy between $M_L$ and $d_L$ as well. 
However, as we have shown, these degeneracies can be broken in the wave-optics regime.

\begin{figure}
    \centering
    \includegraphics[width=0.49\textwidth]{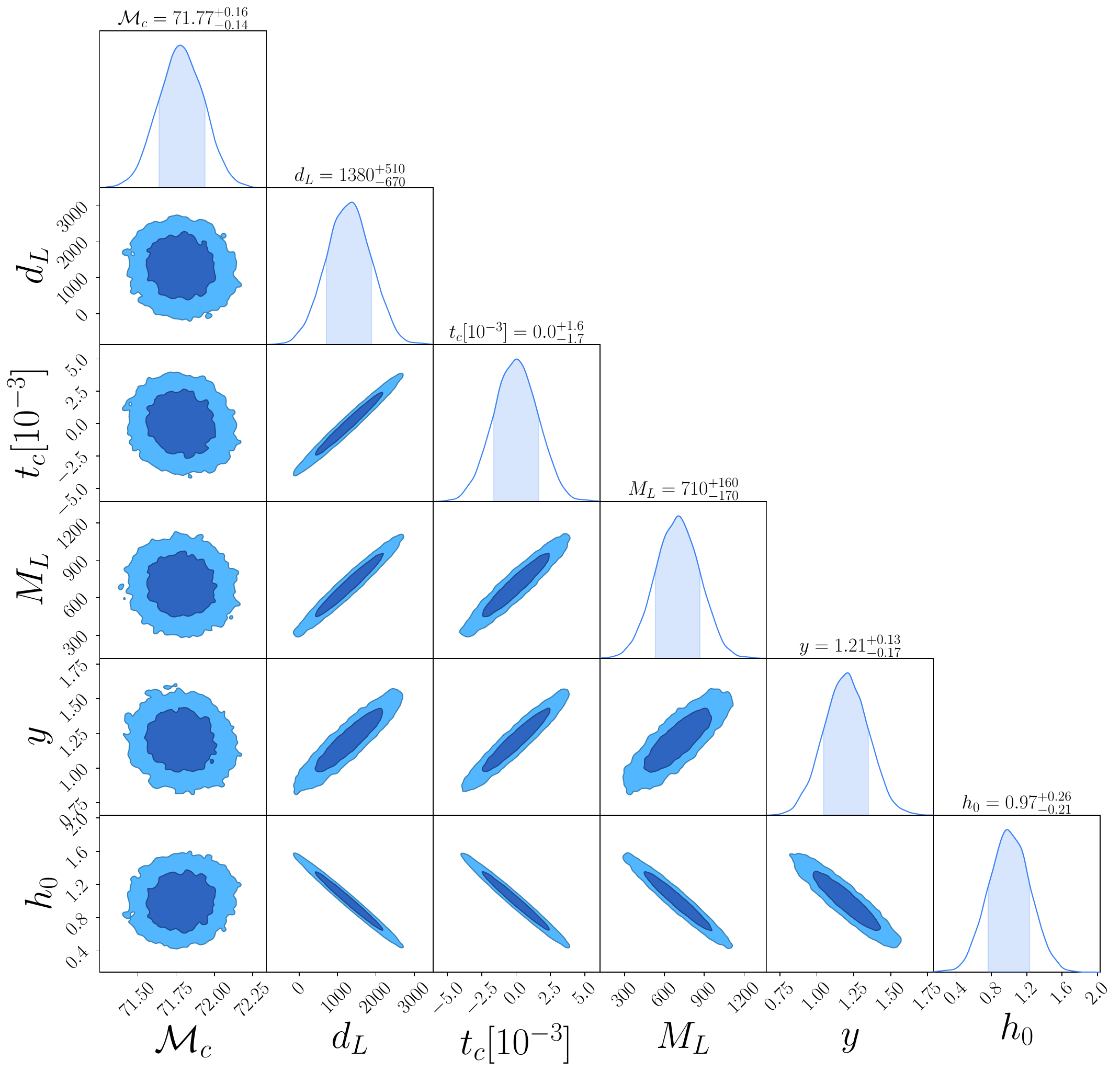}
    \caption{Fisher matrix forecast for constraints on $h_0$ and other parameters as in Figure \ref{fig:fisher_noalign}. The same injected parameters as listed in Table \ref{tab:injpar} are used. The injected value for $h_0$ is $h_0=1$. }
    \label{fig:h0_Fisher}
\end{figure}

By fixing $m=1$ and $\kappa_c=0$, we perform Fisher matrix forecasts for constraints on $h_0$ under the PM model in the wave-optics regime, along with waveform parameters and lensing parameters, as shown in Figure \ref{fig:h0_Fisher}. The fiducial value for $H_0$ is $H_0^{\rm fid}=70~{\rm km~s^{-1}~Mpc^{-1}}$, and the injected value of $h_0$ is $h_0=1$. Other injected parameters are the same as in the first column of Table \ref{tab:injpar}. We can see that strong degeneracy occurs between $h_0$ and $\{d_L,t_c,M_L,y\}$. The correlation between $h_0$ and $d_L$ is heavily dependent on the relation of $d_L\sim1/H_0$, so that the degeneracy is stronger than for $\lambda-d_L$ in the MSD case. Meanwhile, the correlation between $h_0$ and other parameters is similar to the MSD case.

\subsection{Wave-optics cosmography}
\label{sec:WO_cosmography}

As we have just seen, $H_0$ can cause strong degeneracies in PE for lensed GW signals. However, this also provides a new approach to constrain $H_0$ with wave-optics lensing. By performing PE on lensed waveforms that includes $h_0$ as a free parameter,
we can measure $H_0$ from the posterior of $h_0$ using $H_0=h_0H_0^{\rm fid}$. To investigate the precision to constrain $H_0$ with this method, we perform a Fisher matrix forecast with multiple events based on specific assumptions for the GW source population and the lensing rate for microlensing.

We first generate GW events following the Powerlaw + Peak black-hole mass distribution model \cite{Fishbach:2020ryj,Farah:2021qom,KAGRA:2021duu}, as was also done in the LVK cosmology paper \cite{LIGOScientific:2021aug}. We adopt the mass model parameters as $\alpha=3.78$, $\beta=0.81$, $m_{\rm min}^{\rm BH}=4.98M_\odot$, $m_{\rm max}^{\rm BH}=112.5M_\odot$, $\delta_m=4.8M_\odot$, $\mu_g=32.27M_\odot$, $\sigma_g=3.88M_\odot$ and $\lambda_p=0.03$, based on detected LVK events \cite{KAGRA:2021duu}. We use the \texttt{GWCOSMO} package \cite{PhysRevD.101.122001,10.1093/mnras/stac366,Gray:2023wgj} to generate the Powerlaw + Peak mass distribution, which is shown in Figure \ref{fig:mass_distribution} in Appendix \ref{sec:app_cosmo}.

Following Ref. \cite{Callister:2020arv}, we use the Madau-Dickinson star formation rate redshift evolution function \cite{Madau:2014bja} to describe the redshift distribution of the binary merger rate, which gives 
\begin{equation}
     {\cal R}(z)= R_0 [1+(1+z_p)^{-\gamma-k}] \frac{(1+z)^\gamma}{1+[(1+z)/(1+z_p)]^{\gamma+k}}.
     \label{eq:madau_rate}
\end{equation}
We adopt the same value $R_0=30~{\rm yr}^{-1}{\rm Gpc}^{-1}$ as in \cite{Callister:2020arv}, which is consistent with the GWTC-3 population study \cite{KAGRA:2021duu}.
We also adopt $\gamma=4.59$, $k=2.86$ and $z_p=2.47$ for the Madau-Dickinson model as in \cite{LIGOScientific:2021aug}. 

Next, we calculate the rate of microlensing for GW events in terms of the optical depth. We assume that the lens mass density is constant in comoving volume and takes up $10\%$ of the matter density of the universe (This is an arbitrary choice, as this work only provides a proof of principle). 
The differential optical depth is given in \cite{Vietri_1983,Zumalacarregui:2017qqd} as 
\begin{equation}
    \frac{d\tau}{dz_L} = \frac{3}{2} \alpha \, \Omega_{m,0} H_0^2 \frac{(1+z_L)^2}{H(z_L)}\frac{D_LD_{LS}}{D_S},
    \label{eq:optical_depth}
\end{equation}
where $\alpha=0.1$ is the fraction of matter density contributing to lens density, and we adopt the 2018 Planck TT+lensing+BAO value $\Omega_{m,0}=0.3065$ \cite{planck_2018}. By integrating Eq. \eqref{eq:optical_depth} over $z_L$ up to the source redshift, we obtain the optical depth of lensing effects for GW sources, which is plotted in Figure \ref{fig:optical_depth} in Appendix \ref{sec:app_cosmo}. The lensed GW event rate as a function of redshift is then calculated by multiplying the merger rate redshift evolution function with the optical depth in terms of redshift. A plot showing the lensed event rate compared with the binary merger rate is presented in Figure \ref{fig:event_number} in Appendix \ref{sec:app_cosmo}.
Note that the fraction of dark matter made of these lenses does not affect the redshift dependence of the optical depth and only rescales the overall rate. 
This fraction should be chosen not to conflict with current constraints on massive compact halo objects (e.g. \cite{Mroz:2024wag}). 
In our case, we are interested mostly in simulating lensed events and not in their relative rate compared to non-lensed events.

\begin{figure}
    \centering
    \includegraphics[width=0.49\textwidth]{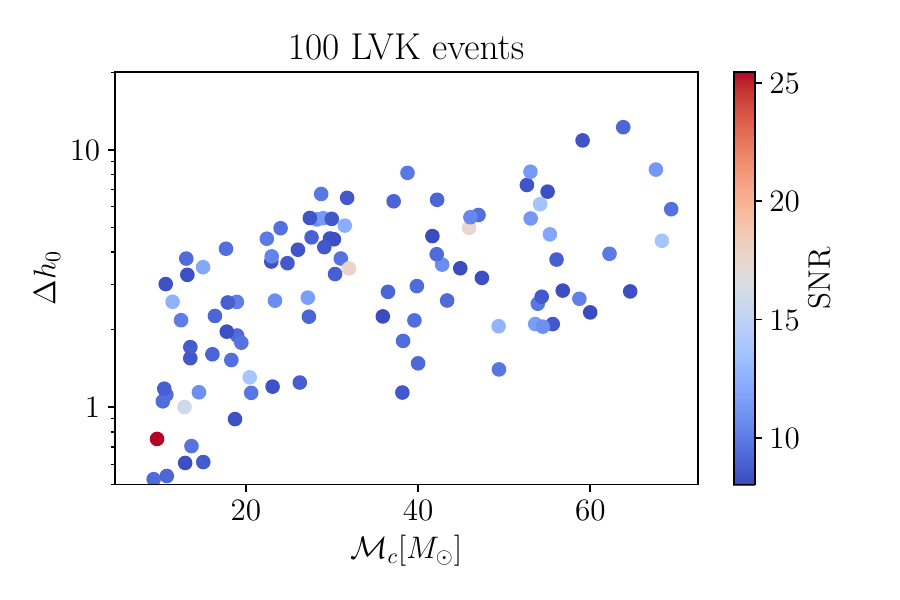}
    \caption{The plot shows 100 simulated lensed GW events in the wave-optics regime detected by a ground-based GW detector with LIGO A+ design sensitivity \cite{aLIGO_design_curve}. The lens mass is assumed to be $100\,M_\odot$ for each event.
    The horizontal axis shows the redshifted chirp mass of the simulated events. The vertical axis shows the uncertainty in $h_0$ for each event. The colour scale shows the SNR of each event. 
    }
    \label{fig:h0_error}
\end{figure}

For the next step, we draw lensed GW events with random masses and redshifts with the probability weighted by the mass distribution and the lensing rate probability distribution computed above. 
The redshift range is $z<1$ since LVK sources at high redshift are likely to have SNR lower than the detection threshold (unless strongly lensed). 
We select events with SNR $>8$. Then we make a simple assumption that each lens weights $100\,M_\odot$, because we want to ensure that the lensed GW events are in the wave-optics regime, since LVK BBH event total masses are around 10 to 100 solar masses. And we want to focus on analysing the factors from the sources that affect constraints on $h_0$ by fixing lens masses. We will also investigate the effect of changing lens masses later. In addition, we assume that the lenses are distributed uniformly in comoving volume, which gives
\begin{equation}
    p(z_L) \propto \frac{dV_c}{dz_L} = \frac{c^3}{H_0^3}\frac{1}{E(z_L)}\bigg(\int_0^{z_L} \frac{dz'}{E(z)'}\bigg)^2,
\end{equation}
where $E(z)=\sqrt{\Omega_{m,0}(1+z)^3+\Omega_{\Lambda,0}}$, and \mbox{$\Omega_{\Lambda,0}=1-\Omega_{m,0}$}. So we draw the redshift $z_L$ of the corresponding lens for each event randomly from a uniform-in-comoving-volume distribution below the source redshift. 
In addition, we assign a random value for the impact parameter $y$ from a uniform distribution in $(0,10)$ for each lens. 
Then we construct the lensed waveform under the transformation with $h_0$ with the PM lens model, and forecast the constraints on $h_0$. Given the lens mass allowed to cause wave-optics lensing for LVK events, the lenses are likely stellar-origin or intermediate black holes, so that they can be well described by the PM lens model. We simulate the uncertainties in $h_0$ via FM for 100 lensed events with the LIGO A+ design sensitivity curve \cite{aLIGO_design_curve} for
one LIGO detector,
and show them in Figure \ref{fig:h0_error} along with the redshifted source chirp masses and SNRs.

\begin{figure}
    \centering
    \includegraphics[width=0.49\textwidth]{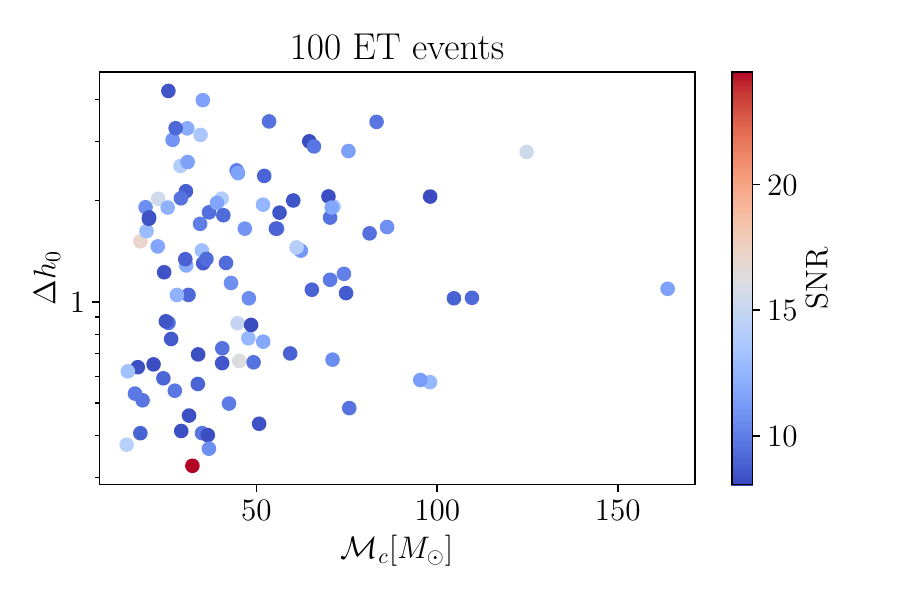}
    \caption{The plot shows 100 simulated lensed GW events in the wave-optics regime similar to Figure \ref{fig:h0_error}, but detected by the Einstein Telescope with sensitivity as in \cite{Hild:2010id}.
    }
    \label{fig:h0_error_ET}
\end{figure}

Figure \ref{fig:h0_error} shows that the uncertainties in $h_0$ from Fisher matrix forecasts
for wave-optics lensed waveforms for PM lenses of $100\,M_\odot$ are mostly on the order of unity or larger. This means that the uncertainty of constraints on $H_0$ is mostly larger than 100\%, which is equivalent to unconstrained results. Such uncertainties are extremely large compared to other $H_0$ measurements, which are currently reaching the percent level precision (see the review in \cite{ABDALLA202249}). 
With LVK detections, we are not expected to have a large enough number of wave-optics lensed events to achieve a precise $H_0$ measurement. However, next-generation (3G) ground-based detectors, such as the Einstein Telescope (ET) \cite{Hild:2010id,ET_design,Maggiore:2019uih} or Cosmic Explorer \cite{Reitze:2019iox,Evans:2021gyd},
are expected to detect significantly more lensed GW events per year \cite{Li:2018stc}. So we also forecast $h_0$ uncertainties with the ET sensitivity curve \cite{Hild:2010id} in Figure \ref{fig:h0_error_ET}. 

From this simulation, we find that the lensed events detected by ET which surpass the SNR threshold of 8 mostly originate from high redshifts within $1<z<4$. This can be explained by the fact that the compact binary merger rate increases at low redshifts, reaches a peak at $z\sim2$, and decreases at higher redshifts. Besides, the lensing rate is monotonically increasing with redshift, so the lensed GW event rate is larger at high redshifts. The uncertainties in $h_0$ obtained with the ET noise curve are lower than for LIGO A+, with many of them lower than $100\%$. 

\begin{figure}
    \centering
    \includegraphics[width=0.49\textwidth]{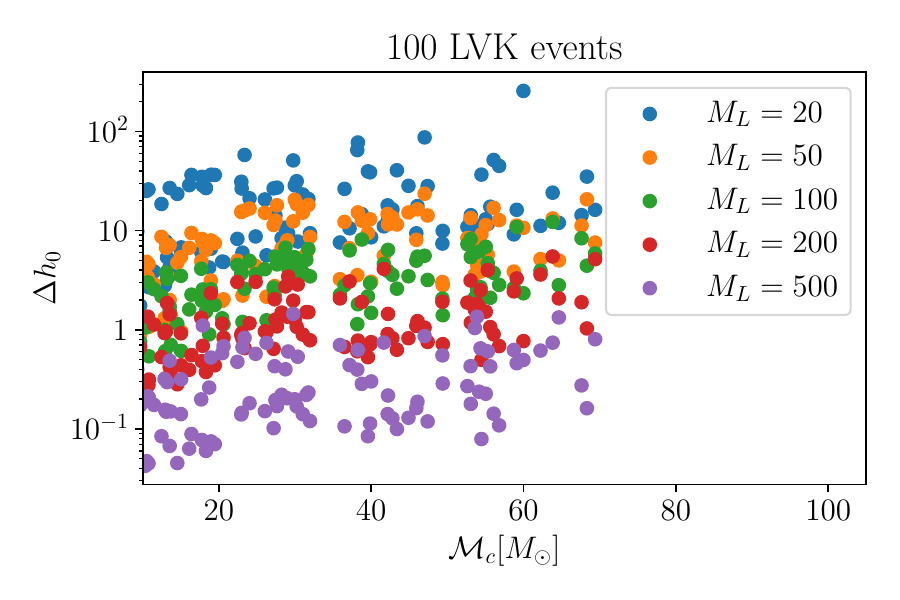}
    \caption{The plot shows the same simulated lensed GW events detected by a LIGO detector as in Fig.~\ref{fig:h0_error}, but with different lens masses shown by different colours.
    }
    \label{fig:h0_error_ML}
\end{figure}

\begin{figure}
    \centering
    \includegraphics[width=0.49\textwidth]{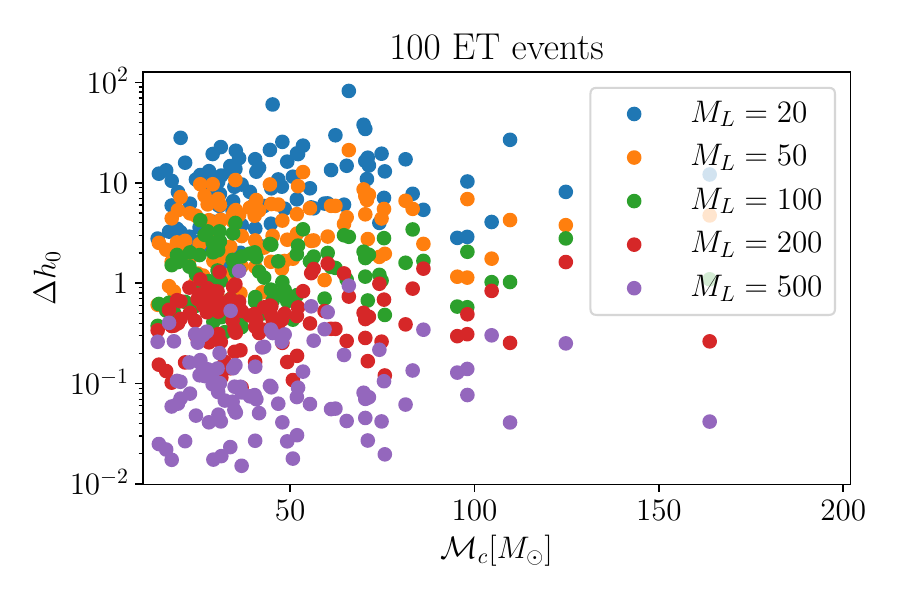}
    \caption{The plot shows the same simulated lensed GW events detected by ET as in Fig.~\ref{fig:h0_error_ET}, but with different lens masses shown by different colours.}
    \label{fig:h0_error_ML_ET}
\end{figure}

From Fig.~\ref{fig:h0_error_ET} we find that SNR is not the dominant factor that determines $\Delta h_0$. As shown in Fig.~\ref{fig:Fisher_error_vary}, the uncertainty on the degeneracy parameter $\lambda$ decreases while lens mass increases. In other words, the ratio of ${\cal M}_c/M_L(1+z_L)$ is the main factor that determines $\Delta \lambda$, and hence $\Delta h_0$ as well. Since $M_L$ is fixed to $100\,M_\odot$ in our population study, events with lower ${\cal M}_c$ give lower $\Delta h_0$, which matches with Figs.~\ref{fig:h0_error} and \ref{fig:h0_error_ET}. We also simulate $\Delta h_0$ for the same lensed GW events with different lens masses and plot them in Fig.~\ref{fig:h0_error_ML} and \ref{fig:h0_error_ML_ET} for A+ LIGO design and ET sensitivity, respectively. As expected, $\Delta h_0$ reduces for higher lens mass. However, as discussed in Sec.~\ref{sec:implications_astro}, the Fisher matrix forecast for the degeneracy parameter cannot represent the real uncertainties for lens masses significantly higher than the source mass. Therefore, the actual uncertainties for $h_0$ with high lens masses could be larger than what we see in Figure \ref{fig:h0_error_ML} and \ref{fig:h0_error_ML_ET}.


In summary, this forecast presents a possibility to measure $H_0$ with wave-optics lensed GW events by 3G detectors. By combining constraints on $h_0$ from multiple lensed events, the uncertainty can drop to a much lower level. However, readers should be aware that our forecast constraints are based on our assumption of a simple lens population model and a fixed lens mass, so our forecasts may not represent the realistic precision for future detections. The current work mainly stands as a proof of principle. We will leave a more accurate forecast with a more realistic lens population model and lensed GW detection rate for 3G detectors to future works.

\section{Conclusions}
\label{sec:conclusions}

In this work, we have investigated the invariance transformations in wave-optics GW lensing. These transformations, the most well-known of which is the mass-sheet degeneracy (MSD), could bias the information that we get from lensed GWs, if not properly accounted for. 

In the first part of the paper (Secs.~\ref{sec:lensing} and \ref{sec:transformations}), we presented a detailed review of gravitational lensing and the invariance transformations, how to derive the MSD and the important equations to study it in GW lensing, e.g. how the amplification factor is impacted by the MSD parameter $\lambda$.

We studied the practical impacts of the MSD, in Sec.~\ref{sec:methods}, with three different methods with increasing complexity and computational expense: template mismatch, Fisher matrix (FM), and Bayesian parameter estimation (PE). We used these methods on a limited set of injections and presented here a couple of the most characteristic cases.
All three analyses agree very well with each other (see e.g. Fig.~\ref{fig:mismatch_ML_lambda} for mismatch and Fig.~\ref{fig:PEvsFM_PM} for FM and PE). 
In particular, they highlighted the degeneracy between the lens parameters, $M_L$ and $y$, that characterize the MSD. Moreover, the FM and PE analyses showed how, with the introduction in the analysis of the MSD parameter $\lambda$, the correlation between the mass of the lens and the impact parameter is shifted to $M_L-\lambda$ and $d_L-\lambda$. The consequence of introducing a new parameter in the analysis is that the uncertainty of the inferred parameters will be larger. This also means, though, that these estimations are more realistic and not biased by the degeneracy.

These results are expanded and treated more in detail in Secs/~\ref{sec:implications_astro} and \ref{sec:implications_cosmo}, where we analyzed the implications of such transformations on astrophysical and cosmological studies. 
In particular, regarding astrophysical conclusions from GW observations, we showed how not considering the MSD could lead to biased values of the parameters that most correlate with $\lambda$ (see Fig.~\ref{fig:values_recap}). However, once we add the MSD parameter to the analysis, the expanded uncertainties encompass the correct values.
The overall impact of this additional parameter on the uncertainty budget is shown in Fig.~\ref{fig:errors_recap}, along with a discussion on how the degeneracy could complicate the reconstruction of the lens model. 

As for the cosmology studies, we analyzed the degeneracy caused by the change of the scaled Hubble constant relative to the fiducial value, $h_0$, which is degenerate with the lens mass due to the invariance transformation, and with the luminosity distance of the source. We then performed a population study by simulating lensed GW events following a LVK source mass distribution model, and assuming a lens density uniform in comoving volume and fixed lens mass of $100\,M_\odot$. We used the Fisher matrix to forecast the constraints on $h_0$ at LIGO A+ sensitivity and ET sensitivity. 
While the uncertainties in $h_0$ are over 100\% for most of the events with current sensitivity, many of those with next-generation sensitivity (such as ET) reduce to the order of $10\%$. 
Better constraints could be achieved for larger lens masses. 
Our forecast reveals a possibility to measure the Hubble constant with wave-optics lensed GW events.  
A more accurate forecast needs to be done with a better lens population model and it will be worth also exploring future space-based detectors like LISA \cite{amaroseoane:2017} that could detect very high SNR events.

Summarizing, we have described the invariance transformations in GW lensing, with a stress on the MSD. We showed how, in the wave-optics regime, this can be partially broken at the expense of a higher uncertainty on some of the inferred parameters. Therefore, we can say that in an ideal case the results we can get are promising. Once we consider more realistic cases and population studies, though, the situation gets more complicated and the uncertainties can become very large, making it very hard to extract valuable information, like $H_0$ measurements, from such events.
Future GW detectors may still yield more promising results.

%%%%%%%%%%%%%%%%%%%% ACKNOWLEDGMENTS %%%%%%%%%%%%%%%%%%
\section*{Acknowledgments}

We thank Otto Hannuksela and Jason Poon for useful discussions on the topic in this work.
AC is supported by a PhD grant from the Chinese Scholarship Council (grant no.202008060014), and the STSM GravNet grant during his visit to the NBI. 
PC and DK are supported by
the Universitat de les Illes Balears (UIB);
the Spanish Agencia Estatal de Investigaci{\'o}n grants
CNS2022-135440,
PID2022-138626NB-I00,
RED2022-134204-E,
RED2022-134411-T,
funded by MICIU/AEI/10.13039/501100011033,
the European Union NextGenerationEU/PRTR,
and the ERDF/EU;
and the Comunitat Aut{\`o}noma de les Illes Balears
through the Direcci{\'o} General de Recerca, Innovaci{\'o} I Transformaci{\'o} Digital
with funds from the Tourist Stay Tax Law
(PDR2020/11 - ITS2017-006)
as well as through the Conselleria d'Economia, Hisenda i Innovaci{\'o}
with grant numbers
SINCO2022/6719
(European Union NextGenerationEU/PRTR-C17.I1)
and SINCO2022/18146
(co-financed by the European Union
and FEDER Operational Program 2021-2027 of the Balearic Islands).
JME is supported by the European Union’s Horizon 2020 research and innovation program under the Marie Sklodowska-Curie grant agreement No. 847523 INTERACTIONS, and by VILLUM FONDEN (grant no. 53101 and 37766).
PC and DK thankfully acknowledge the computer resources at Picasso and the technical support provided by Barcelona Supercomputing Center (BSC) through grants
AECT-2021-3-0014,
AECT-2022-1-0024,
AECT-2022-2-0028,
AECT-2022-3-0024,
AECT-2023-1-0023,
AECT-2023-2-0032,
AECT-2023-3-0020,
AECT-2024-2-0028
from the Red Espa{\~n}ola Supercomputaci{\'o}n (RES). They thank the Supercomputing and Bioinnovation Center (SCBI) of the University of Malaga for their provision of computational resources and technical support.
This paper has been assigned document numbers
\href{https://dcc.ligo.org/\dcc}{LIGO-\dcc}
and \tds.

%%%%%%%%%%%%%%%%% APPENDICES %%%%%%%%%%%%%%%%%%%%%

\appendix

\section{The second minimum in mismatch}

\begin{figure}
    \centering
    \includegraphics[width=0.49\textwidth]{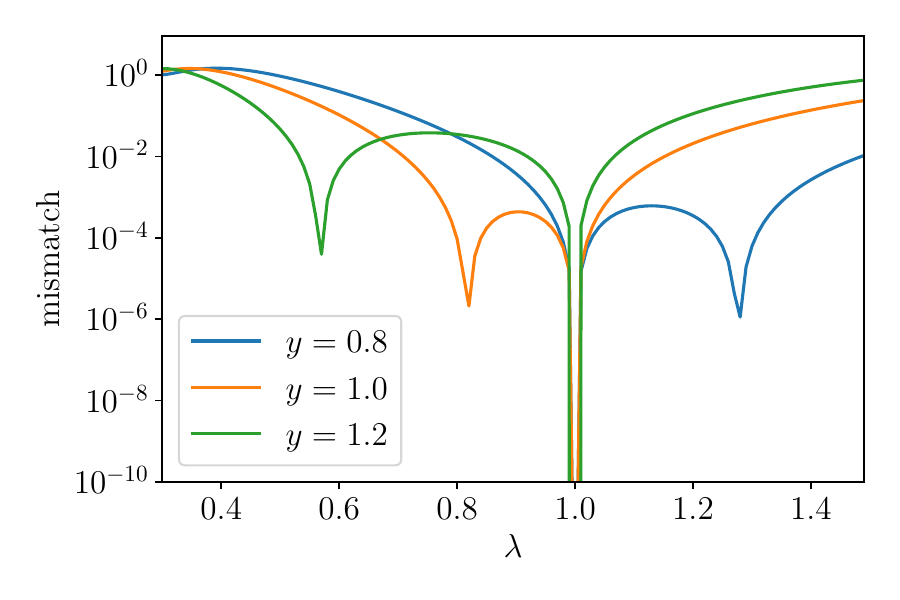}
    \caption{Mismatch along the constant product curve of $M_L\lambda$ with different values of $y$ for the PM model. The waveform parameters and the fiducial lens parameters are the same as in Fig.~\ref{fig:mismatch_ML_lambda}.}
    \label{fig:mismatch_second_min}
\end{figure}

As shown in the lower panel of Figure \ref{fig:mismatch_ML_lambda}, the mismatch computed by aligning the waveforms with the fiducial time delay has a second minimum apart from the fiducial point of $M_L=700,\lambda=1$. It is located along the curve of a constant product $M_L\lambda$, at a lower $\lambda$ value. We plotted the mismatch along this constant product curve in one dimension to examine the second minimum for the PM model, as shown in Figure \ref{fig:mismatch_second_min}. Unlike the minimum at the fiducial value, the second minimum does not drop to 0. We further plotted the mismatch for different values of $y$, and found that the second minimum point varies with $y$. However, it does not vary with any waveform parameters. Moreover, when we plotted the one-dimensional mismatch along the constant product curve of $M_L\lambda$ for the SIS model, there is no second minimum, for different values of $y$. Therefore this feature of a second minimum is specific to the PM lens model. This feature is also present in Fig.~7 of \cite{Cremonese:2021puh} as a second peak of the ratio between SNR and the optimal SNR.

\section{Parameter estimation analyses}
\label{ap:PE}
In this section, we present in more detail the injection and PE cases considered in Sec.~\ref{subsec:PE}. For our Bayesian PE analysis we took into consideration three main cases:
(i) an injection with a PM lens model with a high SNR, divided into two subcases, one with $\lambda=1$ and one with $\lambda=0.8$;
(ii) an injection with a SIS lens model with a high SNR, divided into two subcases, one with $\lambda=1$ and one with $\lambda=0.85$;
(iii) a realistic GW200208-like event, using a PM model.

The injections are done using the \texttt{pycbc} package \cite{pycbc}. The PSD used is the \texttt{aLIGOZeroDetHighPower}\footnote{\href{https://dcc.ligo.org/T070247/public}{dcc.ligo.org/LIGO-T070247/public}.}.
For the injections, we assume detection at three detectors: Hanford, Livingston \cite{LIGOScientific:2014pky} and Virgo \cite{Acernese_2015}.
The amplification factor has been computed in \texttt{python} and \texttt{Mathematica} in the form of a look-up table, following Eqs.~\eqref{eq:F_lambda_PM} and \eqref{eq:F_lambda_SIS} for PM and SIS lens, respectively.
The PE runs have been done using \texttt{bilby 2.2.3} through \texttt{bilby-pipe 1.3.0} \cite{Ashton:2018jfp}.

\subsection{PM injection}
We now explore in more detail the PE for the point mass model with the different simulated events in order to quantify the possible biases induced by the MSD.

\subsubsection*{Injection $\lambda=1$}
\label{subsec:lam1}
For this first case, we will use an injection described by the parameters reported in Table~\ref{tab:injpar}. We choose the source parameters to have an event with a high SNR and the lens parameters such that we are in the wave-optics regime in the relevant frequency band\footnote{
    The condition for which the geometrical optics approximation is broken is given by \cite{Takahashi:2016jom} $M_{3D,L}\leq10^5 M_{\odot}\left[1+z_L)f/\text{Hz}\right]^{-1}$,
where $M_{3D,L}$ and $z_L$ are the 3D mass and redshift of the lens, respectively. This condition is, in general, dependent on $y$, and holds true for $y\sim1$ \cite{Ezquiaga:2020gdt,Bulashenko:2021fes}.
}. 

\paragraph*{PE with fixed $\lambda$}
\label{subsec:lam1-nolam}

\begin{figure}[htbp]
    \centering
    \includegraphics[width=.49\textwidth]{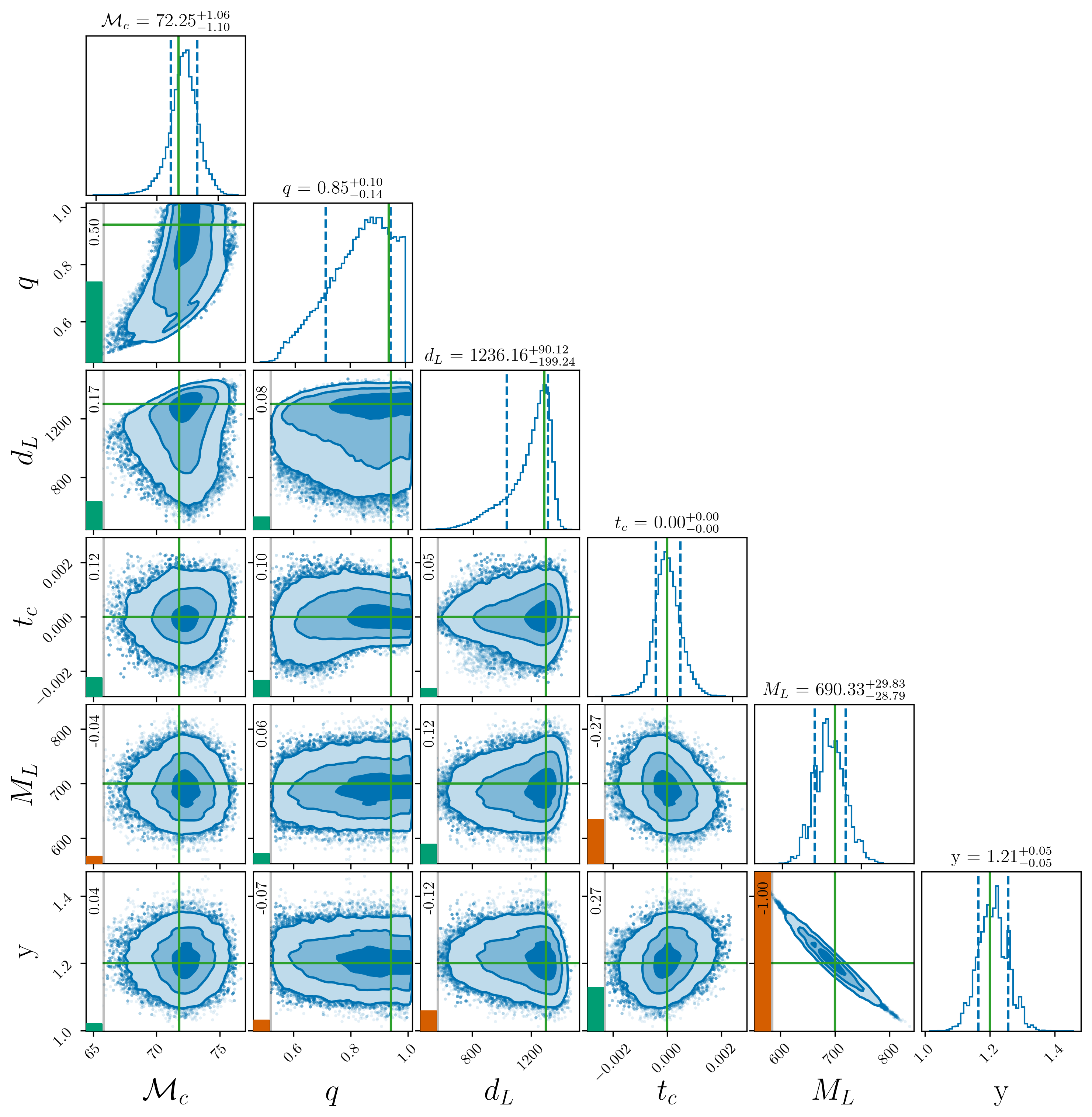}
    \caption{
        Injection with $\lambda=1$ and PE analysis with fixed $\lambda=1$ for the PM model.
    }
    \label{fig:lam1-nolam}
\end{figure}

First, we perform a ``classic'' lensed PE analysis, i.e. an analysis without taking into consideration the MSD.
In Fig.~\ref{fig:lam1-nolam} we show the corner plot of the PE analysis for this case and for the parameters we are interested in, i.e. chirp mass, $\mathcal{M}$; mass ratio, $q$; luminosity distance, $d_L$; cosines of the inclination angle of the source, $\cos\theta_{JN}$; mass of the lens, $M_L$; and impact parameter, $y$. Here, and in the whole paper, we refer to the detector frame parameters, so, for example, $M_L$ stands for Mass of the Lens (measured from the detector frame).

The (anti)correlation\footnote{As we are interested in the absolute value of the correlation, and not on its sign, we will just refer to it as ``correlation'', whether it is positive or negative.} between the lens parameters, $M_L$ and $y$, is $1$. 
Given the MSD, this is exactly what we expect. Also worth noting is that the lens parameters do not correlate significantly with any other parameter. 

As we can see, the uncertainties on the lens parameters are rather low, i.e. $\approx4\%$, and in line with previous results \cite{Cremonese:2021puh}. 
As we shall see, though, these are not the ``real'' uncertainties, since we are not considering the MSD yet.
All the parameters are retrieved correctly without any bias. This is to be expected since the injection with $\lambda=1$ is included in the PE model without $\lambda$, i.e. $\lambda=1$ is the case where no transformation is present and, therefore, we are not yet considering any source of degeneracy.

\paragraph*{PE with free $\lambda$}
\label{subsec:lam1-lam}
This case is treated in the main text with the corner plot shown in (the blue contours of) Fig.~\ref{fig:PEvsFM_PM}.
In this case the $M_L-y$ correlation is translated to $\lambda$ and the uncertainties on the parameters increase. A detailed discussion on this can be found in Sec.~\ref{subsec:PE}.

\subsubsection*{Injection $\lambda=0.8$}
\label{subsec:lam08}

Now, we analyze a case given by an injection with $\lambda\neq1$, in particular $\lambda=0.8$, with all the other parameters left the same as before (see Table~\ref{tab:injpar}). 
This case is of particular interest since a value of $\lambda<1$ mimics a sheet of mass in between the lens and the observer, that could go unseen and bias the inference of the lens parameters.
We aim to see where the degeneracy acts and how a ``transformed'' signal is recovered through the conventional and ``degenerate'' (i.e. adding the $\lambda$ parameter) PE analysis and whether we would be able to recover the parameters of the lens correctly. 
With $\lambda\neq1$, also the lens parameters are ``transformed'', see Eqs.~\eqref{eq:impact_lambda} and \eqref{eq:sigma_lambda}. In particular, we expect that the biased parameters would be $y_\lambda=0.8\cdot1.2=0.96$ and $M_{l,r,\lambda}\approx690$ M$_\odot$.

\begin{figure}[htbp]
    \centering
    \includegraphics[width=.49\textwidth]{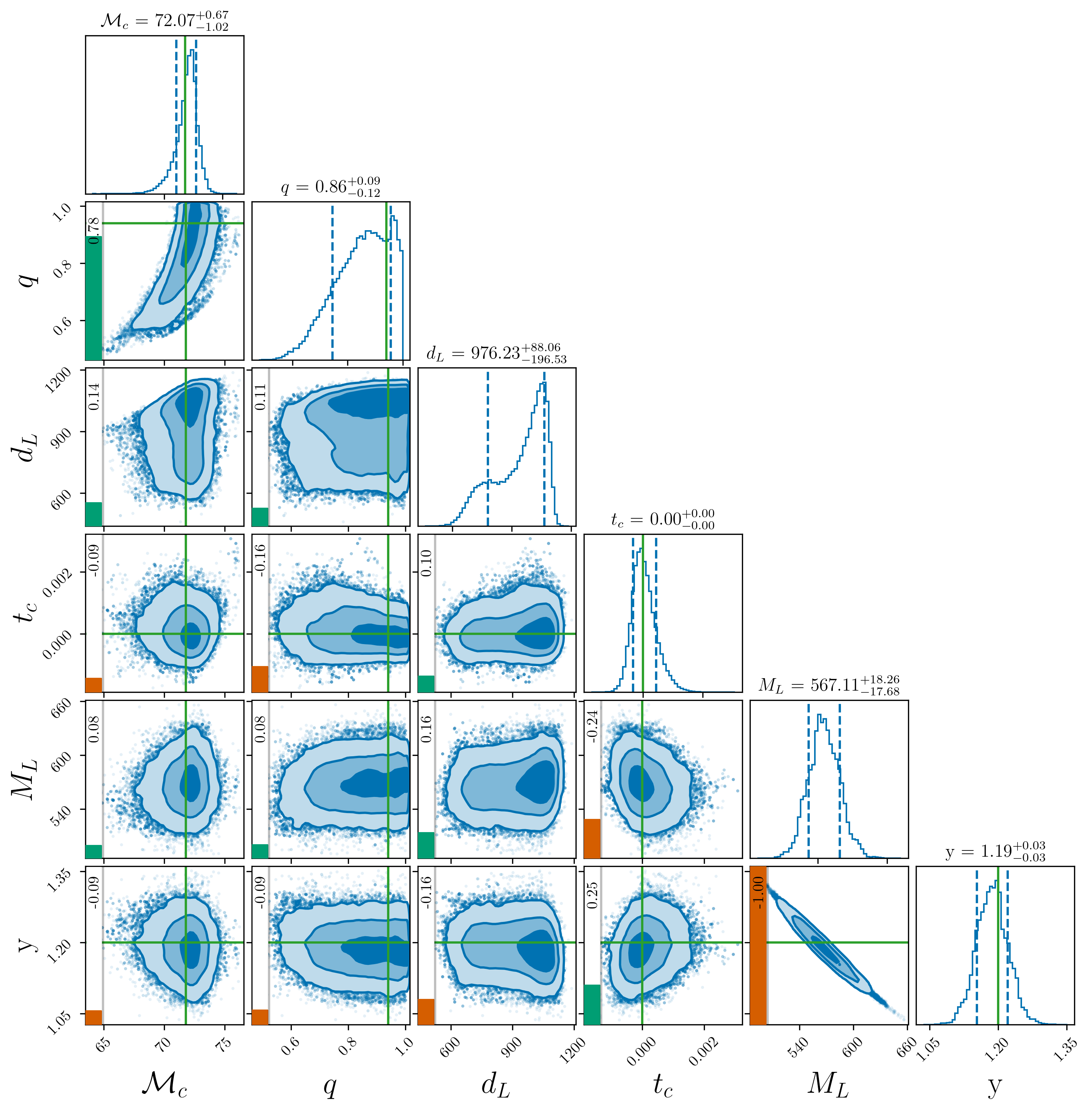}
    \caption{
        Injection with $\lambda=0.8$ and PE analysis with fixed $\lambda=1$ with PM model.
    }
    \label{fig:lam08-nolam}
\end{figure}

\paragraph*{PE with fixed $\lambda$}
\label{subsec:lam08-nolam}
As before, we begin with the case of a ``regular'' PE analysis, with the MSD parameter fixed to $\lambda=1$.
As we can see from Fig.~\ref{fig:lam08-nolam}, the change in the MSD parameter, $\lambda$, is again absorbed by the lens mass. In fact, its value is off by $\approx20\%$ from the real value, while $y$ is retrieved correctly. 
Here, too, we can see that the luminosity distance is affected as well, being off by almost $25\%$. Both the uncertainties, on $M_L$ and $d_L$, are coherent with the first case shown in Fig.~\ref{fig:lam1-nolam}.

An important aspect to notice is that the lens mass retrieved here is not what we expect from the MSD transformation. We are expecting a slightly lower value than the case with $\lambda=1$, but it is significantly lower. We can explain this by looking at the luminosity distance, $d_L$. The distance is lower, which translates into a louder signal, and this would compensate for the ``missing mass'' that we were expecting. 

\begin{figure}[htbp]
    \centering
    \includegraphics[width=.48\textwidth]{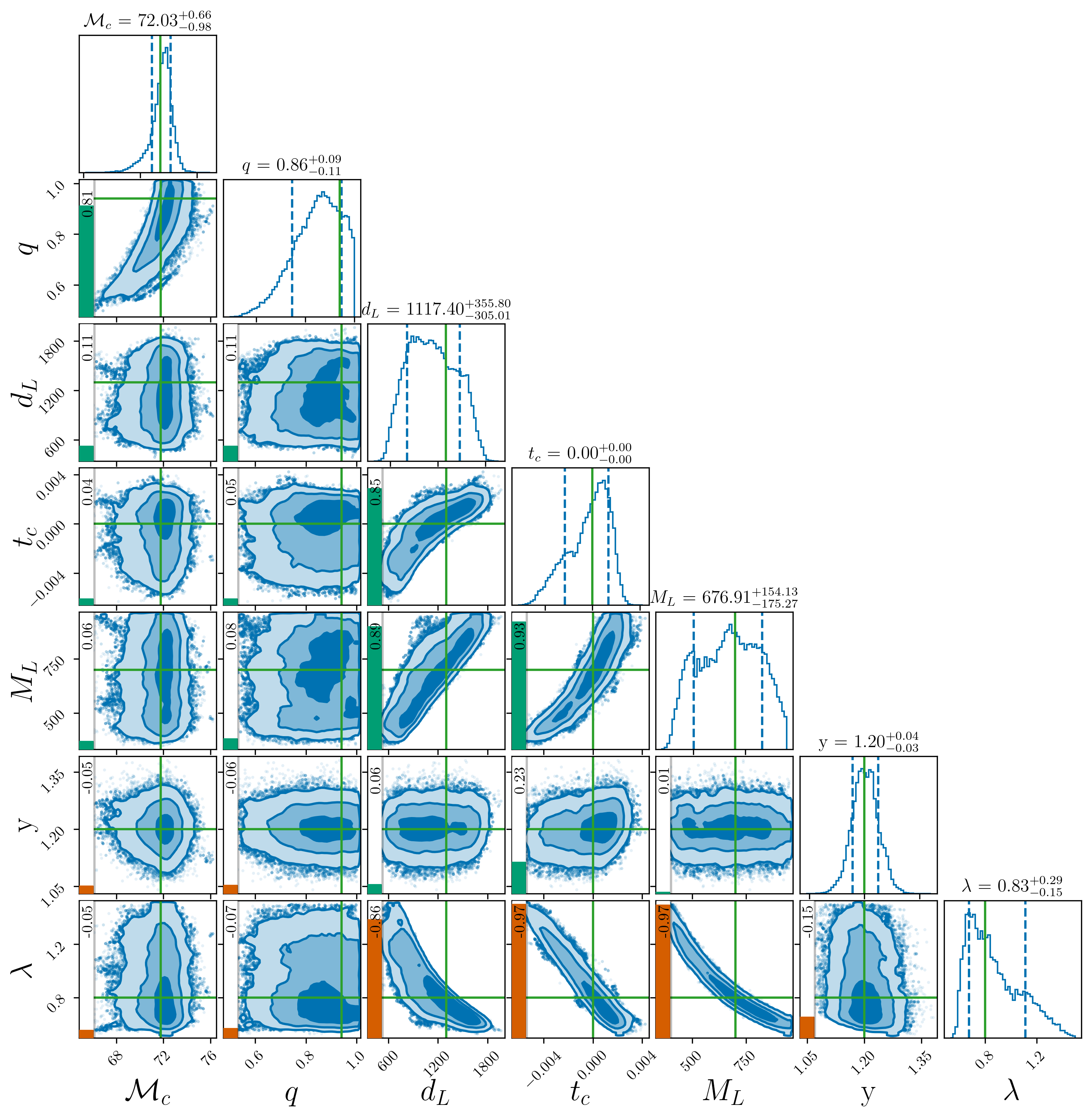}
    \caption{Injection with $\lambda=0.8$ and PE analysis with free $\lambda$.}
    \label{fig:lam08-lam}
\end{figure}

\paragraph*{PE with free $\lambda$}
\label{subsec:lam08-lam}
As before, we now analyze the same event with the addition of $\lambda$ as a free parameter. The result of the PE analysis for this case is shown in Fig.~\ref{fig:lam08-lam}. This confirms what was said above: the degeneracy is mainly ``absorbed" by $M_L$ and $d_L$. These two parameters are retrieved correctly this time, differently from the case above, but the errors are still considerably higher than those with fixed $\lambda$.

\begin{figure}[htbp]
    \centering
    \includegraphics[width=.45\textwidth]{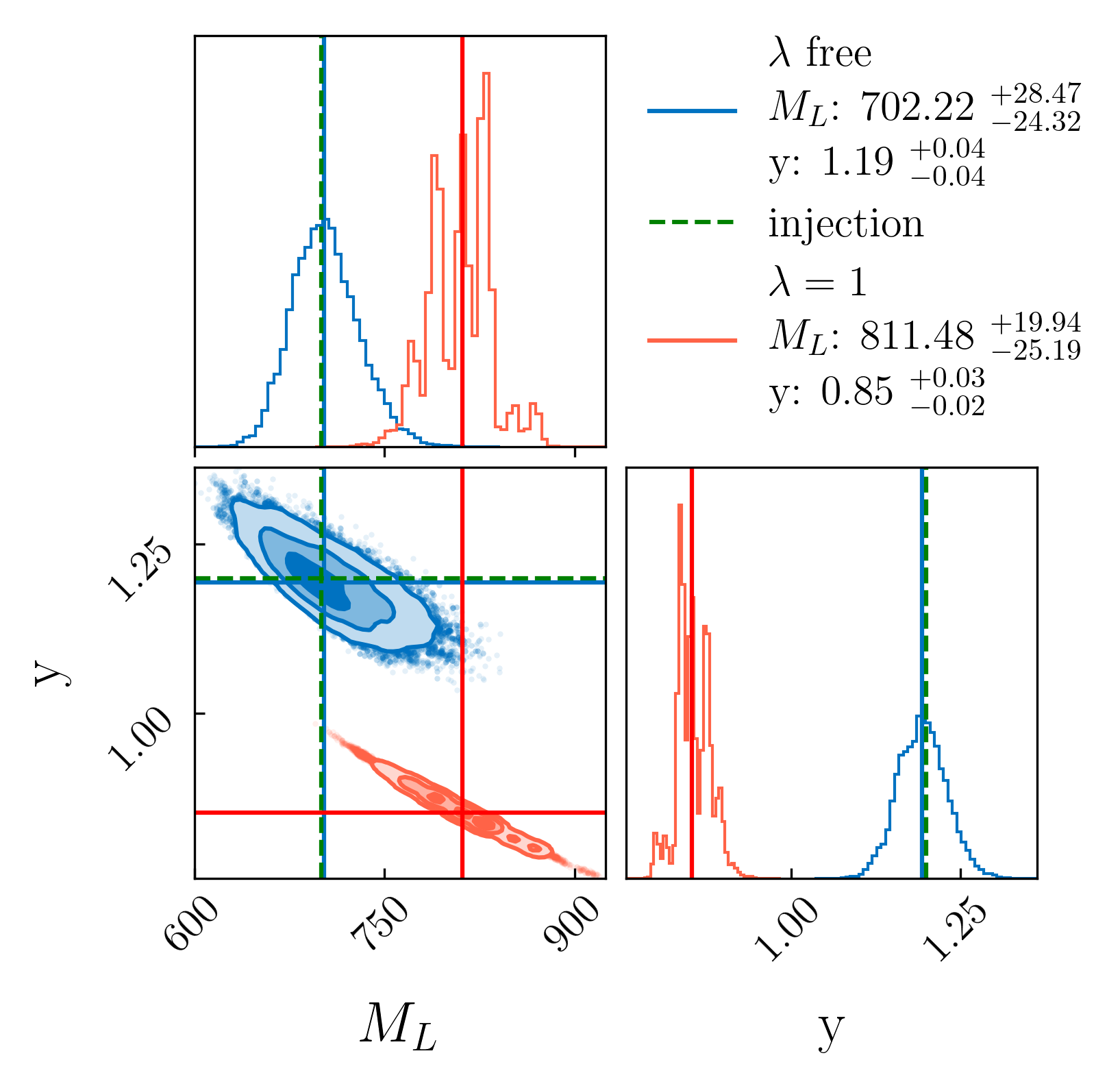}
    \caption{Comparison between injections with $\lambda=0.8$ and PE analysis with free $\lambda$ (blue) and fixed $\lambda=1$ (red). 
    }
    \label{fig:lam-nolam}
\end{figure}

The interesting part of this case is that, even though we added a sheet of mass between the lens and the observer, we can retrieve the values of the parameters of the lens correctly. 
However, this goes at the expense of the precision of the parameters' values, since we are broadening the parameter space. 
In fact, the uncertainties on $M_L$ and $d_L$, which correlate the most with the MSD parameter, $\lambda$, increase considerably w.r.t. the previous case, where these values, though, were wrong.
This is further confirmed by Fig.~\ref{fig:lam-nolam}, where we are comparing two PE analysis with all the main parameters fixed at the injection values, except the lens parameters and $\lambda$ for the blue contours and just the lens parameters for the red contours.

\subsection{SIS injection}

Just like for the PM model case, we also analyze different cases for the SIS model.

\subsubsection*{Injection $\lambda=1$}

\begin{figure}[htbp]
    \centering
    \includegraphics[width=.49\textwidth]{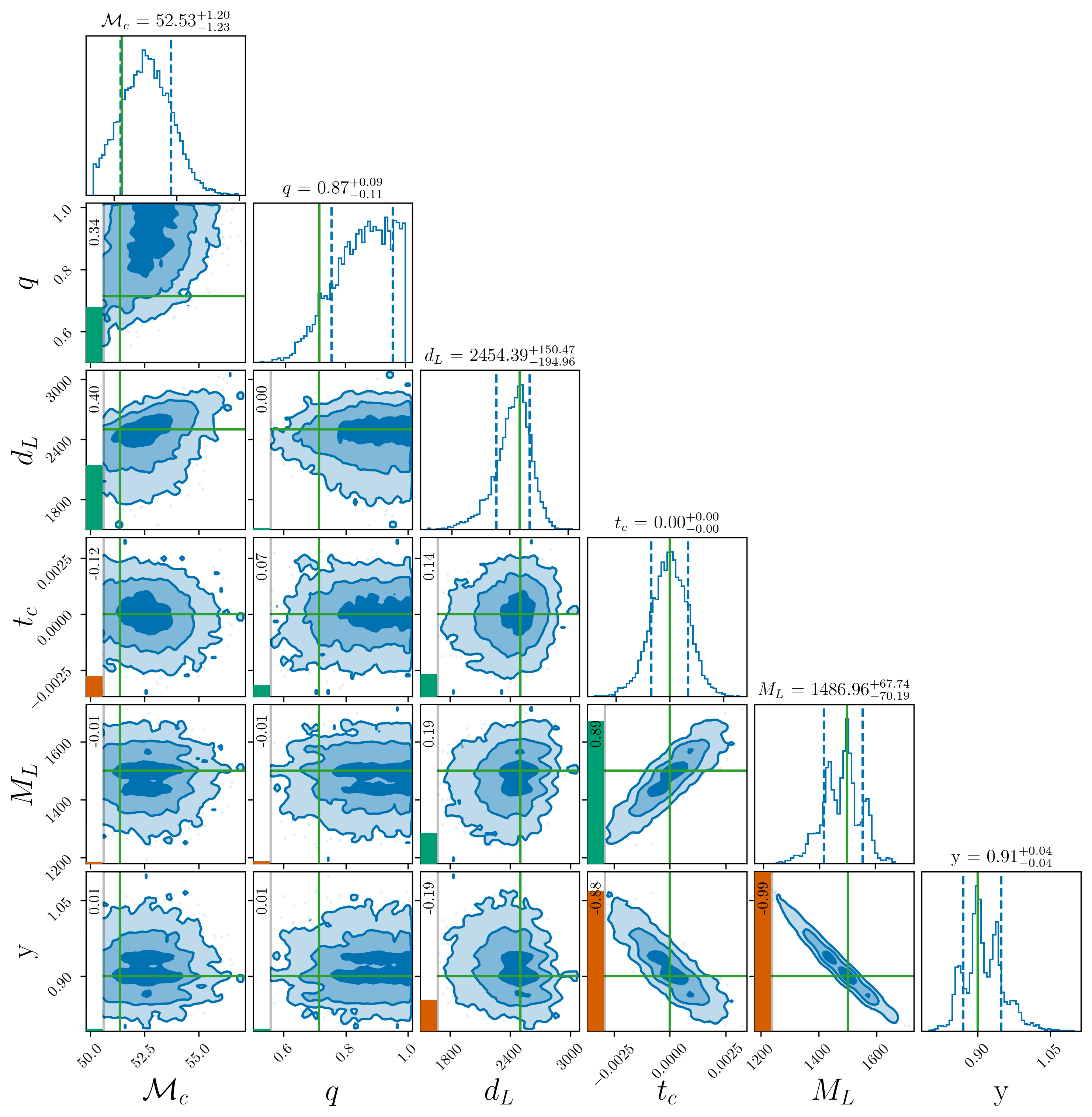}
    \caption{
        Injection with $\lambda=1$ and PE analysis with fixed $\lambda=1$ for the SIS model.
    }
    \label{fig:sis-nolam}
\end{figure}

First of all, we consider a ``standard" injection with $\lambda=1$ (i.e. no transformation). 
Both the cases of PE with fixed and free MSD parameters closely resemble the ones from the PM case. In fact, in Fig.~\ref{fig:sis-nolam}, which shows the case with $\lambda$ fixed, we see a strong correlation between the lens parameters. 

This correlation is translated to the mass of the lens and the distance to the source when we leave $\lambda$ free, as we showed in Fig.~\ref{fig:PEvFM_sis} in the main text. The only small difference is a slight increase in the (low) correlation between the impact parameter, $y$, and $\lambda$ w.r.t. the PM case.

\subsubsection*{Injection $\lambda=0.85$}

\begin{figure}[htbp]
    \centering
    \includegraphics[width=.49\textwidth]{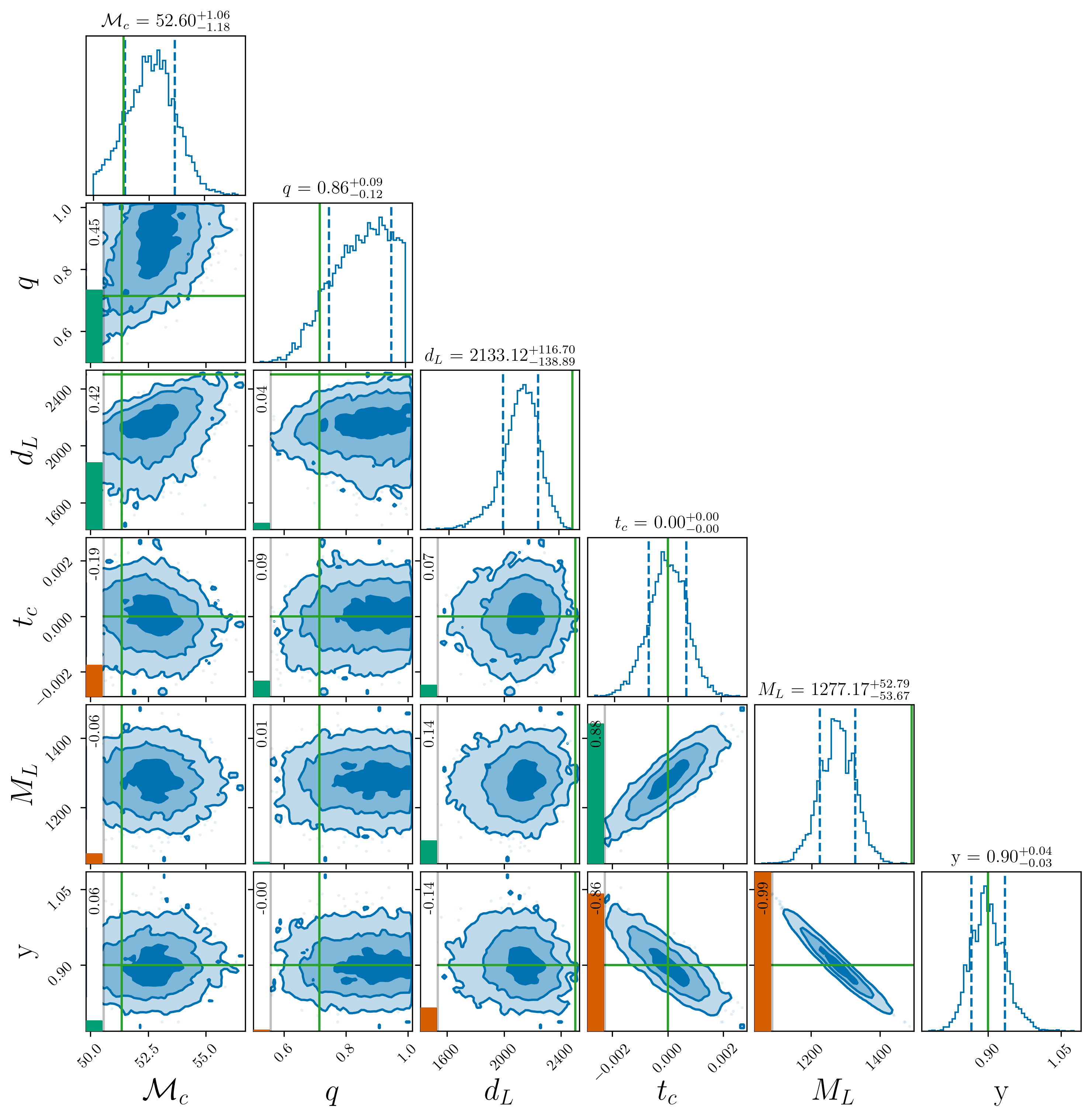}
    \caption{
        Injection with $\lambda=0.85$ and PE analysis with fixed $\lambda=1$ for the SIS model. 
    }
    \label{fig:sislam085-nolam}
\end{figure}

\begin{figure}[htbp]
    \centering
    \includegraphics[width=.49\textwidth]{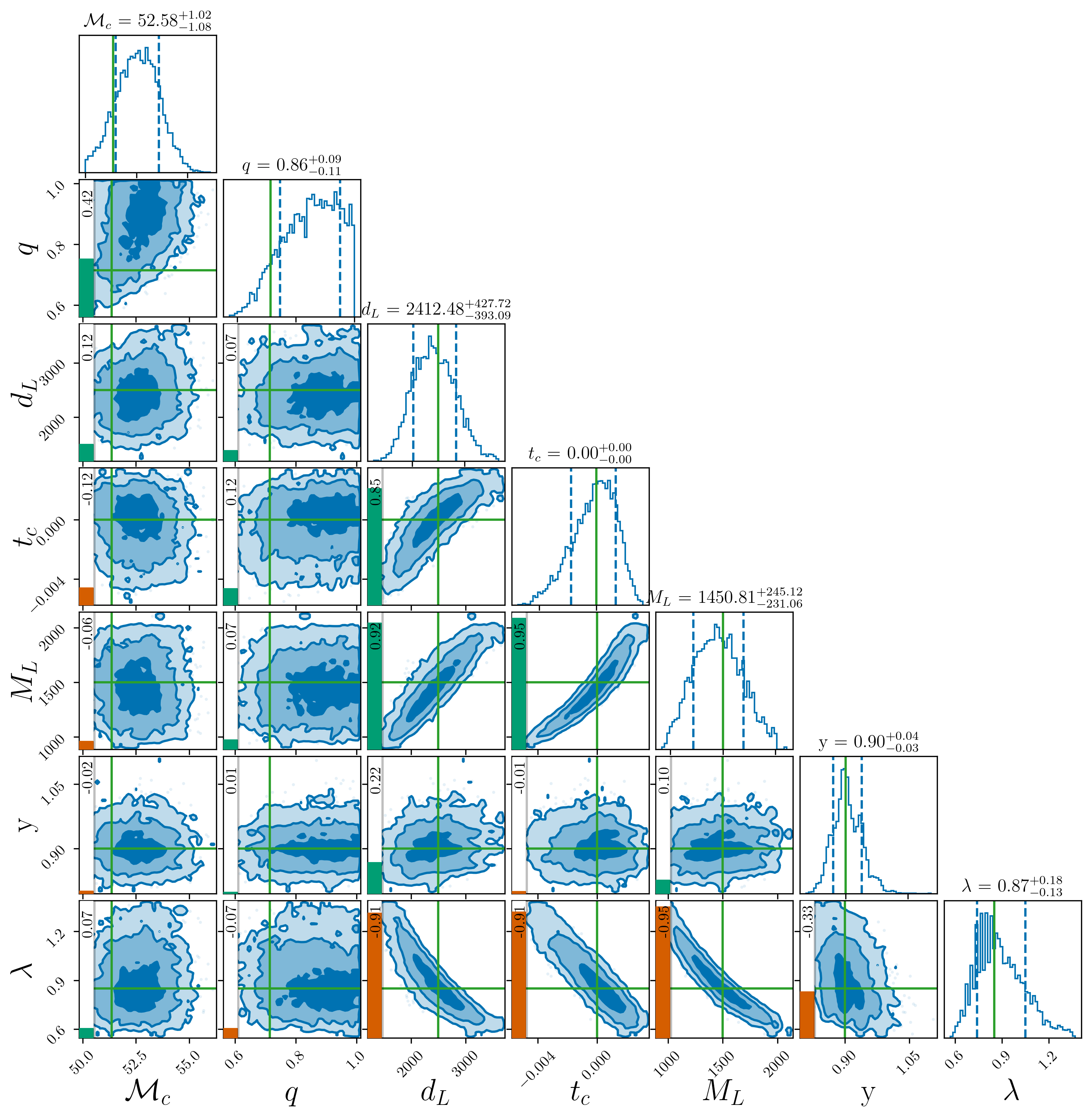}
    \caption{
        Injection with $\lambda=0.85$ and PE analysis with free $\lambda$ for the SIS model. 
    }
    \label{fig:sislam085-lam}
\end{figure}

Figures~\ref{fig:sislam085-nolam} and \ref{fig:sislam085-lam} show the cases of an injection with $\lambda=0.85$ and PE analyses done with the MSD parameter fixed and free, respectively. As was the case for the PM model analyses, here again we find a biased estimation of the lens mass and luminosity distance to the source when we do not consider the MSD (Fig.~\ref{fig:sislam085-nolam}). As expected and shown in Fig.~\ref{fig:sislam085-lam}, leaving $\lambda$ free, on one hand, increases the uncertainty on these two parameters, but, on the other hand, it allows us to retrieve the correct values of the lens mass and the distance.

Overall, the SIS case confirmed what we saw in the previous case, where we considered a PM lens model. Therefore, the observations made there are valid also here.

\subsection{GW200208-like event}

Finally, we present the more realistic case based on the event GW200208. Once again we analyzed the injection both with $\lambda=1$ and $\lambda$ free. The latter case was already presented in the main text, see Fig.~\ref{fig:gw200208-lam}.

\paragraph*{PE with fixed $\lambda$}
\label{subsec:gw200208-nolam}

\begin{figure}[htbp]
    \centering
    \includegraphics[width=.49\textwidth]{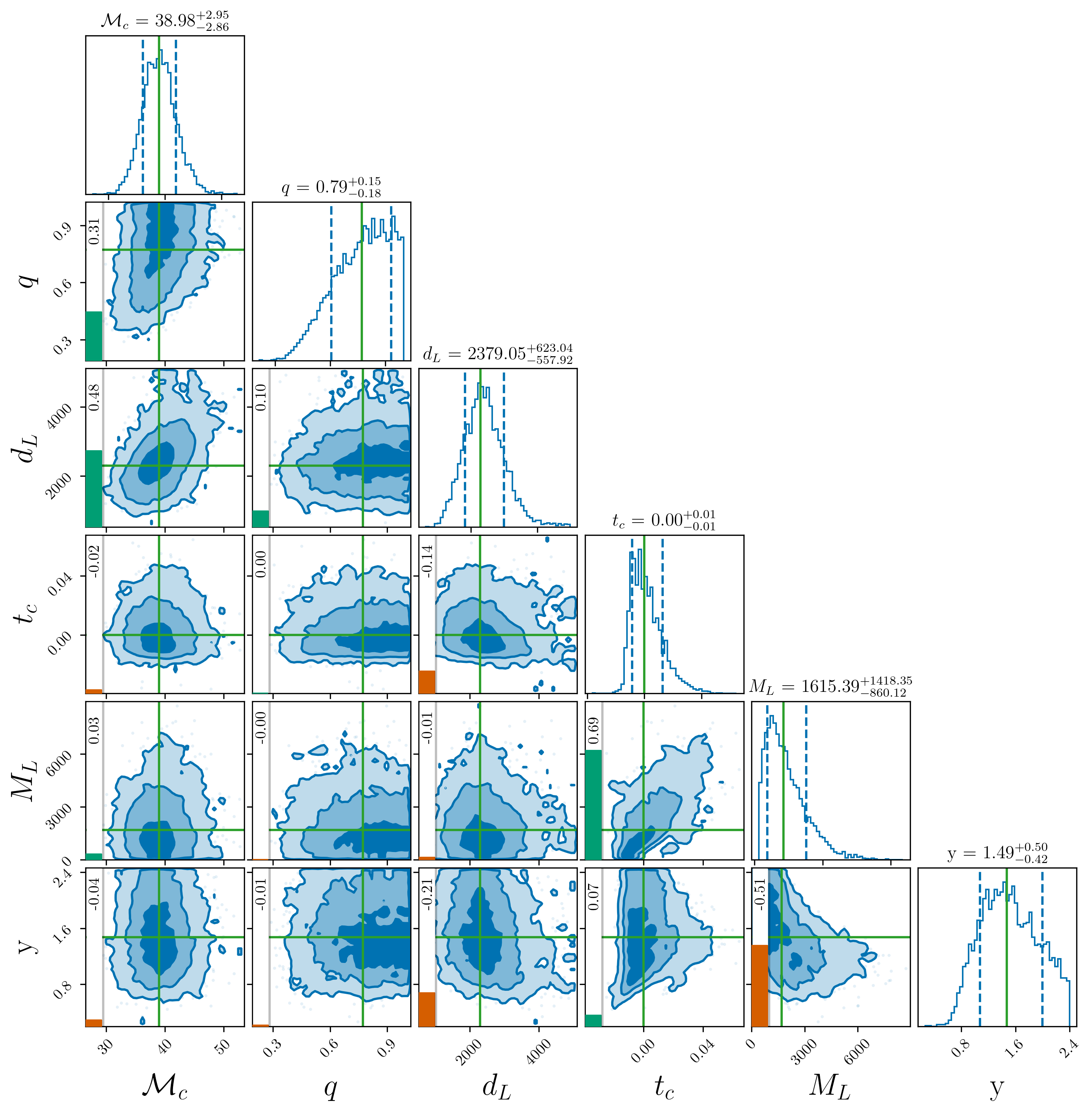}
    \caption{
        Injection with the GW200208-like event and PE analysis with fixed $\lambda=1$, with the PM model.
    }
    \label{fig:gw200208-nolam}
\end{figure}

Here, we set $\lambda=1$ and see the behavior of the ``normal" lensed PE, shown in Fig.~\ref{fig:gw200208-nolam}.
Because of the lower SNR, the posteriors now are not as well defined as before, but we can still see the major feature we saw earlier, i.e. the high correlation between $M_L$ and $y$. 
The values of the lens parameters are slightly biased but in line with the degeneracy that we expect.

\section{Mass distribution and lensing rate in population study}
\label{sec:app_cosmo}

\begin{figure}[t!]
    \centering
    \includegraphics[width=\linewidth]{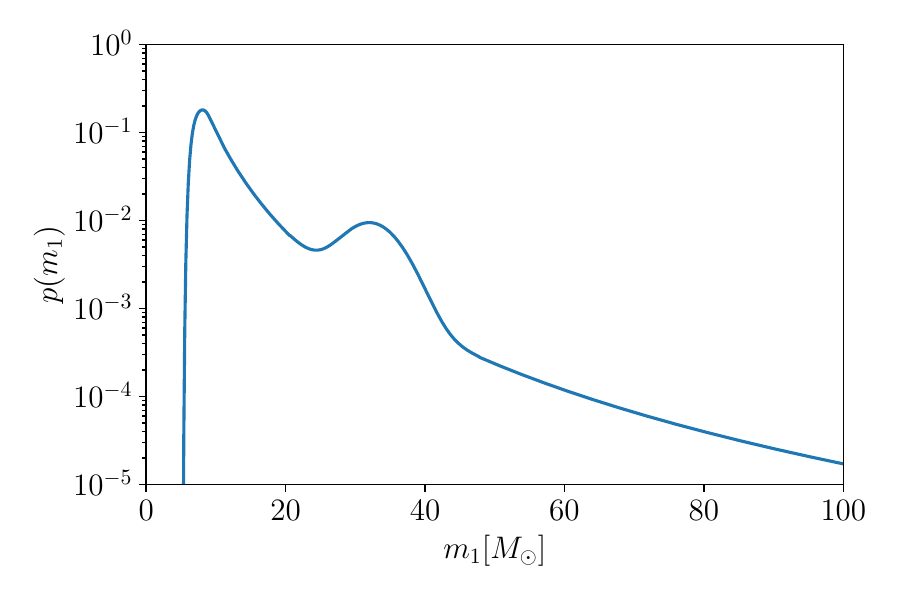}
    \caption{Powerlaw + peak mass distribution model for the GW source population simulation in Section \ref{sec:WO_cosmography}. }
    \label{fig:mass_distribution}
\end{figure}

In this appendix, we present the plots showing the detailed intermediate steps in the population study for wave-optics cosmography discussed in Sec.~\ref{sec:WO_cosmography}. Figure~\ref{fig:mass_distribution} shows the Powerlaw + peak mass distribution model generated with parameters listed in Sec.~\ref{sec:WO_cosmography}, using the \texttt{GWCOSMO} package  \cite{PhysRevD.101.122001,10.1093/mnras/stac366,Gray:2023wgj}.
The component masses of the simulated binary merger events are randomly drawn from this probability distribution, with $m_1>m_2$.

\begin{figure}[t]
    \centering
    \includegraphics[width=\linewidth]{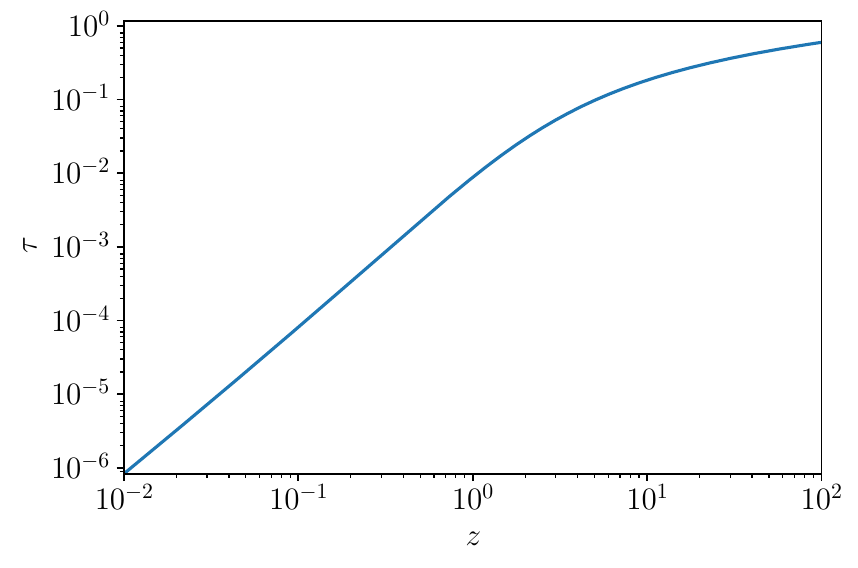}
    \caption{The optical depth by integrating Eq. \eqref{eq:optical_depth} over the lens redshift up to the source redshift denoted as $z$ in the plot. We adopt $\alpha=1$ and $\Omega_{m,0}=0.3065$.}
    \label{fig:optical_depth}
\end{figure}

\begin{figure}[t]
    \centering
    \includegraphics[width=\linewidth]{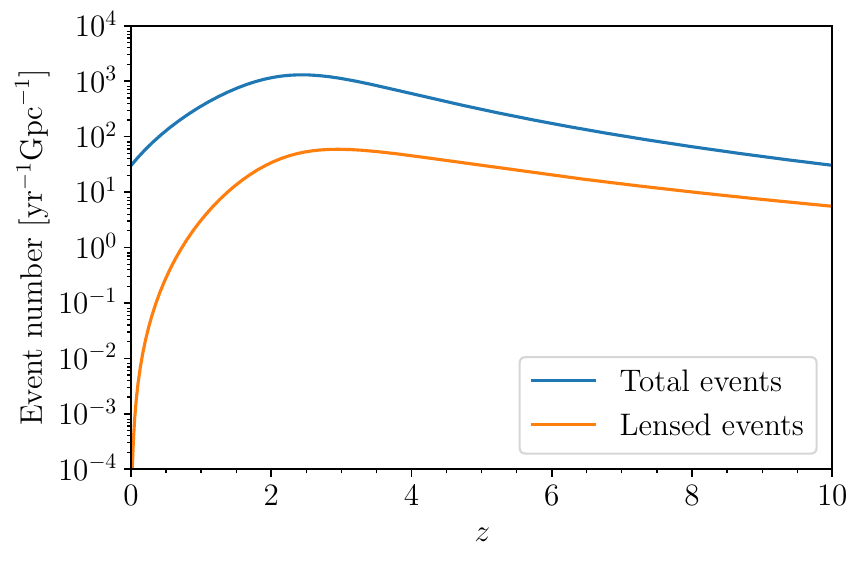}
    \caption{The lensed event rate compared with the total binary merger event rate as a function of redshift, from the calculation in Sec.~\ref{sec:WO_cosmography}. The total merger rate uses the Madau-Dickinson model shown by Eq. \eqref{eq:madau_rate}, with $R_0=30~{\rm yr}^{-1}{\rm Gpc}^{-1}$, $\gamma=4.59$, $k=2.86$ and $z_p=2.47$. The lensed rate is computed by multiplying the total rate by the optical depth shown in Fig.~\ref{fig:optical_depth}.}
    \label{fig:event_number}
\end{figure}

The lensing rate for GW events is computed as the optical depth $\tau$ given the mass density of lenses between the observer and the source. In Sec.~\ref{sec:WO_cosmography}, the optical depth for a source at redshift $z$ is computed by integrating Eq.~\eqref{eq:optical_depth} over the lens redshift from 0 up to $z$, which is shown in Fig.~\ref{fig:optical_depth}. The lensed GW event rate as a function of redshift is then calculated by multiplying the total binary merger event rate with the optical depth in terms of redshift. In Fig.~\ref{fig:event_number} we show the comparison between the total event rate and the lensed event rate. The total binary merger event rate is plotted using the Madau-Dickinson model given by Eq.~\eqref{eq:madau_rate}, with $R_0=30~{\rm yr}^{-1}{\rm Gpc}^{-1}$, $\gamma=4.59$, $k=2.86$ and $z_p=2.47$. But in the population study in Sec.~\ref{sec:WO_cosmography}, the lensed event rate is normalized as a probability distribution function for randomly drawing source redshifts, so the values of $R_0$ and the lens density fraction $\alpha$ are not important in our simulation.

%%%%%%%%%%%%%%%%%%%% REFERENCES %%%%%%%%%%%%%%%%%%

\bibliographystyle{apsrev4-1}
\bibliography{biblio}{}

\end{document}